\documentclass{pasj00}
\usepackage[varg]{txfonts} 
\color{black}
\newcommand{\fluxunit}{erg~cm$^{-2}$~s$^{-1}$}
\newcommand{\sbunit}{erg~cm$^{-2}$~s$^{-1}$~sr$^{-1}$}

\begin{document}
\Received{$\langle$reception date$\rangle$}
\Accepted{$\langle$acception date$\rangle$}
\Published{$\langle$publication date$\rangle$}
\SetRunningHead{H. Akamatsu et al.}{X-Ray Study of the Outer Region of Abell 2142 with Suzaku}
\title{
X-Ray Study of the Outer Region of Abell 2142 with Suzaku
\thanks{Last update: \today}}
\author{
H. Akamatsu\altaffilmark{1},
A. Hoshino\altaffilmark{2},
Y. Ishisaki\altaffilmark{1},
T. Ohashi\altaffilmark{1},
K. Sato\altaffilmark{3},
Y. Takei \altaffilmark{4},
N. Ota \altaffilmark{5}
}
\altaffiltext{1}{
Department of Physics, Tokyo Metropolitan University,\\
 1-1 Minami-Osawa, Hachioji, Tokyo 192-0397}
\altaffiltext{2}{
Graduate School of Natural Science and Technology,\\
 Kanazawa University, Kakuma, Kanazawa, Ishikawa 920-1192}
\altaffiltext{3}{
Department of Physics, Tokyo University of Science, \\
1-3 Kagurazaka, Shinjyuku-ku, Tokyo, 162-8601, Japan
}
\altaffiltext{4}{
Department of High Energy Astrophysics, Institute of Space and 
Astronautical Science,\\ Japan Aerospace Exploration Agency,\\
3-1-1 Yoshinodai, Chuo-ku, Sagamihara, Kanagawa 252-5210
}
\altaffiltext{5}{
Nara Women's University, Kitauoyanishi-machi, Nara, Nara 630-8506
}
\email{h\_aka@phys.se.tmu.ac.jp}
\maketitle

\begin{abstract}
  We observed outer regions of a bright cluster of galaxies A2142
  with Suzaku. Temperature and brightness structures were measured out
  to the virial radius ($r_{200}$) with good sensitivity.  We
  confirmed the temperature drop from 9 keV around the cluster center
  to about 3.5 keV at $r_{200}$, with the density profile well
  approximated by the $\beta$ model with $\beta = 0.85$.  Within $0.4\
  r_{200}$, the entropy profile agrees with $r^{1.1}$, as predicted by
  the accretion shock model.  The entropy slope becomes flatter in the
  outer region and negative around $r_{200}$.  These features suggest
  that the intracluster medium in the outer region is out of thermal
  equilibrium.  Since the relaxation timescale of electron-ion Coulomb
  collision is expected to be longer than the elapsed time after shock
  heating at $r_{200}$, one plausible reason of the low entropy is the
  low electron temperature compared to that of ions.  Other possible
  explanations would be gas clumpiness, turbulence and bulk motions of
  ICM\@.  We also searched for a warm-hot intergalactic medium around
  $r_{200}$ and set an upper limit on the oxygen line intensity.
  Assuming a line-of-sight depth of 2 Mpc and oxygen abundance of 0.1
  solar, the upper limit of an overdensity is calculated to be 280 or
  380, depending on the foreground assumption.
\end{abstract}

\section{Introduction}
The Cold Dark Matter scenario of the cosmic structure formation
predicts that clusters of galaxies are formed via collisions and
mergers of smaller groups and clusters.  As shown by numerical
simulations, merging plays a critical role in the cluster evolution.
X-ray observations have provided many pieces of evidences for cluster
mergers, through imaging infalling subclusters and disturbed,
irregular morphologies of intracluster medium (ICM) (e.g.,
\cite{forman82}).  Furthermore, even in apparently relaxed clusters,
discontinuous ICM structures so-called ``cold fronts'' are sometimes
found from the Chandra observations. The cold fronts are interpreted
as contact discontinuity caused by the subcluster collisions
\citep{markevitch07}. Although the morphological data are
accumulating, the physical states of the ICM, particularly the
ionization equilibrium state and the kinematics, are yet to be
clarified.

In the study of the dynamical evolution of clusters, we will focus on
the cluster outer regions since they are expected to contain important
information on the formation process. The cluster outskirts are
connected to the surrounding large-scale structure, where the gas is
falling towards the cluster potential and possibly subject to 
shock heating. Once disturbed by the subcluster collisions, it should
take long time for the gas to settle due to the low density and large
spatial size. Therefore, the outskirts of merging clusters offer us
opportunities to look into the gas in its transition to the thermal
and ionization equilibriums.

Recent X-ray studies with Suzaku showed temperature structure of ICM
to the virial radius ($r_{200}$) for several relaxed clusters
\citep{george08,reiprich09,bautz09,hoshino10,simionescu11}.  
The results show a systematic drop of temperature by a factor of $\sim 3$
from the center to $r_{200}$.  
Also, the entropy shows a flattening or a small drop, after a monotonous increase with radius, 
around the outermost region.  
It has been interpreted as the deviation between electron and ion temperatures in this region.  
Numerical simulations for relaxed clusters also suggest that the outskirts of clusters are
not in hydrostatic equilibrium (e.g.\ \cite{burns10}).  

Compared with the relaxed clusters, however, there is little
information about the ICM properties in the outer regions of merging
clusters. In those dynamically-young systems, the ICM are thought to
be in the early stage of thermal relaxation. Thus, the X-ray
observations of merging clusters and comparisons with the relaxed
clusters will give us new, valuable information on the ICM properties
at large radii and help us to draw the global picture of formation and
evolution of clusters.

The surface brightness around the cluster virial radius is much lower
than the central region (typically, by a factor of $10^4$).  Thus, the
detailed estimation of foreground and background emission as well as
careful assessment of all the systematic errors is important in this
kind of study. The XIS instrument on Suzaku has a low
and stable background and the behavior of the non-X-ray background (NXB)
is well known \citep{tawa08, koyama07}, which makes XIS the most suitable for
the study of the cluster outer regions.

An additional science at cluster outer
regions is the search for the warm-hot intergalactic medium (WHIM)\@.
The filamentary structure of the local universe has been probed
mainly through galaxy distribution (e.g.\ \cite{eisenstein05}).  
Its detailed structure would be directly observed by the WHIM, which has
temperature of $10^6-10^7$ K and contains more than half of baryons in
the local universe (e.g.\ \cite{cen06}).  
Because of its extreme faintness, detailed observations of the WHIM will be the subject of future
high-resolution X-ray studies.  
The gas density and chemical composition of the WHIM are still poorly
known both theoretically and observationally.  
The cluster outskirts are the regions where the 
ICM is connected to the WHIM and will enable us to place observational
constraints about the WHIM properties. 
As shown, by e.g., \citet{takei08}, 
Suzaku XIS is able to set strong constraints about the redshifted oxygen emission.

Abell 2142 (A2142, $z=0.0909$) is a bright cluster of galaxies, having
a high ICM temperature of $kT \approx 9$~keV.  This object is also
known as the first cluster in which the cold fronts have been detected
\citep{markevitch00}.  The cold fronts in the south and the northwest are $\timeform{0.7'}$ (or 70~kpc) 
and $\timeform{2.7'}$ (or 270~kpc) off of the cluster center, respectively, 
and a sharp surface-brightness drop by a factor of about 2 is
seen. Since the temperature and density distributions suggest that the
pressure is constant across the cold front, it is considered as a contact discontinuity. 
The presence of these structures naturally
indicates that A2142 is a merger and the subcluster infall seems to be
occurring along the northwest--southeast direction. 
Actually the overall cluster emission is elongated in this direction. 
This is likely to coincide with the large-scale structure, 
and the matter density seems to be enhanced along the filament.

A2142 is also suitable for the search of emission lines from
the WHIM. Given the cluster redshift of $\sim 0.1$, the OVII line (the rest-frame energy of 0.65~keV) is shifted to
0.57~keV and then falls into a gap between the Galactic OVII line energy 
and the instrumental line features. Therefore, it makes it possible to
distinguish the WHIM emission from the local one.  In addition,
because the redshift of A2142 is not too high, the oxygen line from
the WHIM can be measured with good sensitivity.
With these purposes, we have carried out Suzaku observations of the northwest offset regions of A2142.

In this paper, we will use $H_0 = 70$ km s$^{-1}$ Mpc$^{-1}$, $\Omega_{\rm M}=0.27$
and $\Omega_\Lambda = 0.73$.
This cosmology leads to 100.4 kpc per arcmin at $z=0.0909$.
 The virial radius approximated by $r_{200} = 2.77 h_{70}^{-1}
(\langle T\rangle /10 {\rm keV})^{1/2}{\rm Mpc} /E(z)$, 
with $E(z)=(\Omega_{\rm M}(1+z)^{3}+1-\Omega_{\rm M})^{1/2}$  \citep{henry09}.
For our cosmology and redshift, $r_{200}$ is 2.48 Mpc ($= \timeform{24.8'}$) with $kT = 8.7$ keV\@.
In this paper, we adopted the solar abundance defined by \citet{anders89} and 
the Galactic $N_{\rm H}$ by \citet{dickey90}.
The errors are in the 90\% confidence for a single parameter.

\section{Observations and Data Reduction}
\label{sec:obs}

\begin{table*}[bt]
\caption{
Log of Suzaku observations of Abell 2142
}\label{tab:obslog}
\begin{center}
\begin{tabular}{cccccccccc}
\hline\hline
Position(Obs.\,ID) & Start &End &Exp.\,time (ks)$^*$ &Exp.\,time (ks)$^\dagger$  \\ \hline
Center  (801055010)  &	2007 Jun 04	&2007 Jun 05	&51.4	&45.2\\
Offset1 (802030010)  & 2007 Oct 04	&2007 Oct 05	&37.6	&26.6\\
Offset2 (802031010)  & 2007 Sep 15	&2007 Sep 17	&57.6	&41.3\\
Offset3 (802032010)  & 2007 Oct 29	&2007 Oct 30	&23.7	&20.4\\
T CrB (401043010) &2006 Sep 06	&2006 Sep 08	&46.3	&36.8\\
\hline\hline 
\scriptsize $\ast$:COR2  $>$ 0 GV & \scriptsize $\dagger$ COR2 $>$ 8 GV \\
\end{tabular}
\end{center}
\end{table*}

\begin{figure*}[ht]
\begin{tabular}{cc}
\begin{minipage}{0.5\hsize}
(a)
\\[-1.3cm]
\begin{center}
\includegraphics[scale=0.4]{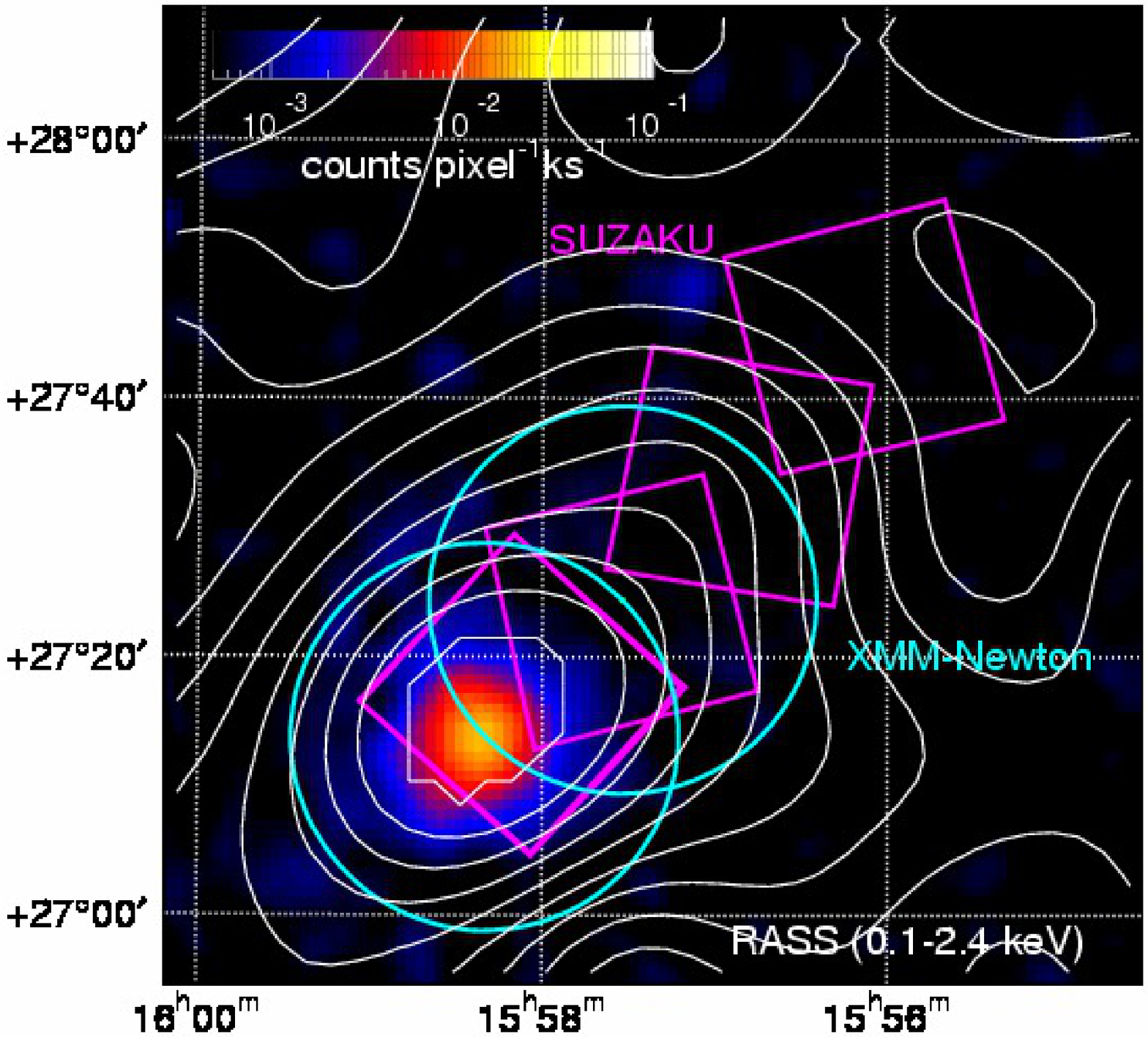}
\end{center}
\end{minipage}
\begin{minipage}{0.5\hsize}
(b)
\\[-1.3cm]
\begin{center}
 \includegraphics[scale=0.4]{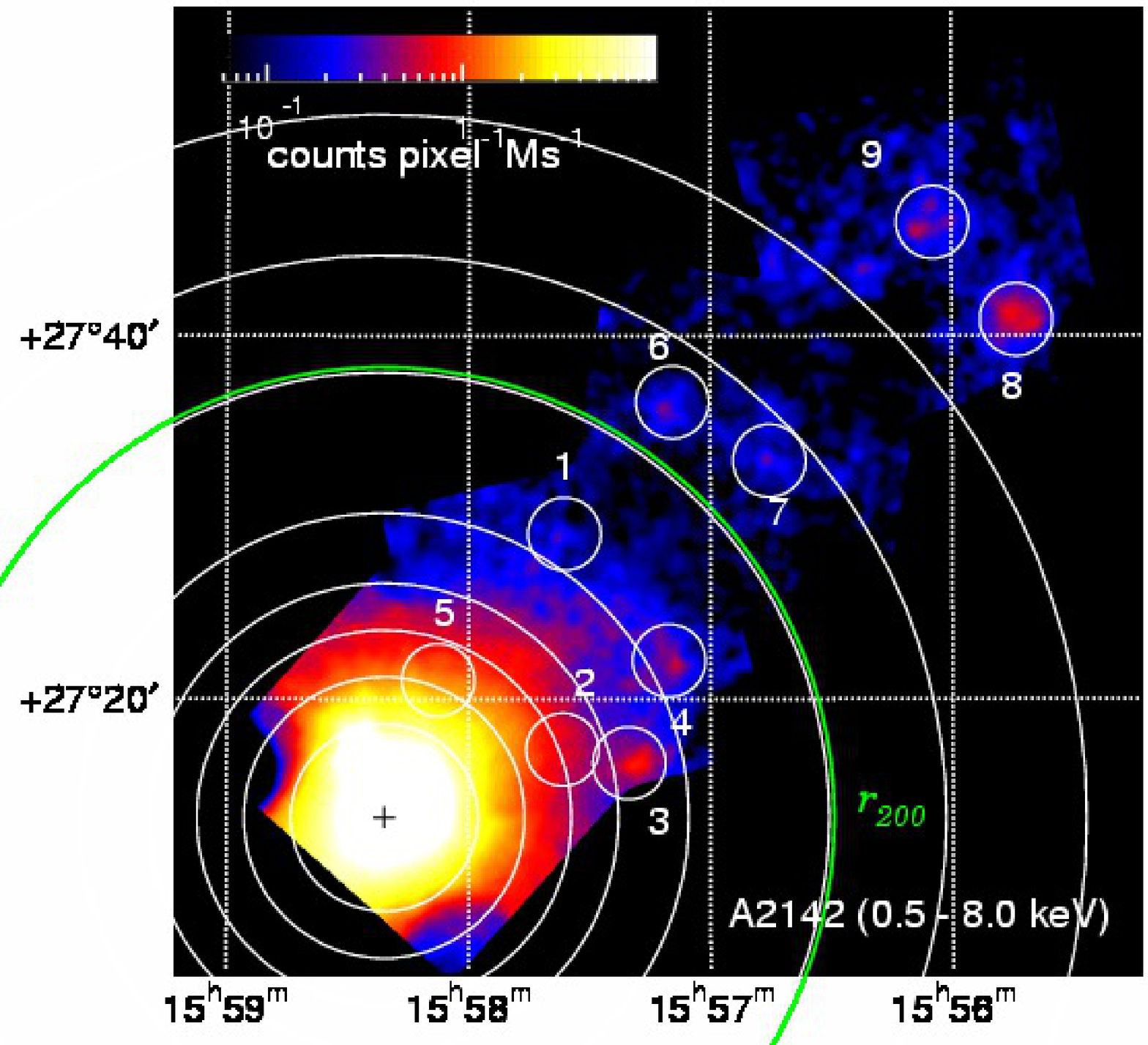}
\end{center}
\end{minipage}
\end{tabular}
\caption{(a):Rosat All Sky Survey image around A2142. Cyan circles show XMM-Newton
  FOV, and Magenta boxes show Suzaku Observations.  
  White contours show galaxy distribution associated with the A2142 taken SDSS catalogue.
  (b):NXB subtracted Suzaku FI+BI image of A2142 in 0.5-8.0 keV band smoothed by a
  2-dimensional gaussian with $\sigma =16$ pixel =$\timeform{17''}$ . 
  The image is corrected for exposure time but not for vignetting.  
  Large white circles indicate the regions used for spectrum analysis. 
  Large green circle show the virial radius of A2142 ($\sim2.5$~Mpc).  
  Small white circles show the excluded point sources.}
\label{fig:Image}
\end{figure*}

As shown in figure~\ref{fig:Image}, we performed four pointing observations
in 2007 with the XIS instrument around Abell 2142. 
The central pointing observation was performed in June, and the other
three 
in September
to October. The cold front feature indicates that there is an
ongoing merger
in the north-west to south-east direction.
This suggests that
matter would be falling in along this merger axis, possibly from a
large-scale filament. We allocated the observed
regions to be successively offset towards the north-west direction.
The regions are designated as Center, Offset1, Offset2, and Offset3
with the exposure times 51.4 ks, 37 ks, 58 ks, and 24 ks,
respectively. The observation log is shown in table~\ref{tab:obslog}.
The outermost observation reaches twice the virial radius
($\timeform{49.2'} \sim 4.92$ Mpc) from the cluster center, 
in which we planned to search for emission from the warm-hot intergalactic medium (WHIM).

Three out of the four CCD chips were available in these observations: XIS0, XIS1 and XIS3. 
The XIS1 is a back-illuminated chip with high sensitivity in the soft X-ray energy range.  
The effective area has been siginificantly reduced by the time of the
observations
due to the contamination building up on
the IR/UV blocking filters.
This effect along with its uncertainty are included in the effective area in our analysis.  
We used HEAsoft ver 6.9 and CALDB 2010-12-06.  All the XIS sensors were in the 
normal clocking mode, and the spaced-row charge injection (SCI) was applied.

We extracted pulse-height spectra in 9 annular regions with boundaries
at 
$\timeform{0'}$, $\timeform{2.5'}$, $\timeform{5.0'}$, $\timeform{7.8'}$, $\timeform{10.3'}$, 
$\timeform{12.9'}$, $\timeform{16.8'}$, $\timeform{24.6'}$, $\timeform{31.1'}$, $\timeform{38.8'}$
centered on ($\timeform{15h58m16s.13}$, $\timeform{27D13'28''}$) from all the XIS events.
We analyzed the spectra in the 0.5--10~keV range for the FI detectors and 0.35--8~keV for the BI detector.  
In all annuli, the calibration source positions were masked out using the {\it calmask} calibration database (CALDB) file.

\section{Background Analysis}
\label{sec:bgd}
In the present study of the ICM emission in the cluster outskirts,
correct estimation of the background is of utmost importance.  
As a standard practice, we assume three background components: 
non-X-ray background (NXB), cosmic X-ray background (CXB) and Galactic emission.  
The Galactic emission component has a spatially variable spectrum.
We estimated the Galactic background spectra using two Suzaku
observation data, one was Offset3 and another one was observation of
TCrB (Observation ID = 401043010) which was located at 1 degree south of A2142.
We will show each background component in this section.

\subsection{Cosmic X-ray Background}\label{sec:cxb}
Since the CXB consists of many extragalactic
point sources, we tried to remove the sources 
as many as possible and then modeled 
the remaining emission with a power-law.
We estimated the CXB surface brightness 
to be  5.97 $\times 10^{-8} $ \sbunit~
after the source subtraction,
 based on ASCA GIS measurements \citep{kushino02}.
We carefully subtracted point sources brighter than $8 \times10^{-14}$
erg cm$^{-2}$ s$^{-1}$, while
the flux limit of the point sources 
eliminated in \citet{kushino02}
is $2\times10^{-13}$ erg cm$^{-2}$ s$^{-1}$.
Our flux limit is thus sufficiently lower than \citet{kushino02}.
The details of point-source subtraction
are described in Appendix~\ref{sec:appa}.

 To estimate the amplitude of the CXB fluctuation,
we scaled the fluctuation measured with Ginga \citep{hayashida89} to
our flux limit and field of view, following the analysis by \citet{hoshino10}.  
The fluctuation amplitude scales as $({\Omega_{\rm e,Suzaku}}/{\Omega_{\rm e,Ginga}} )^{-0.5}$, 
with $\Omega_{\rm e,Suzaku}$ and $\Omega_{\rm e,Ginga}$ the effective
field of views (FOVs) of Suzaku and Ginga instruments, respectively.  We show the
resultant relative fluctuation $\sigma$/$I_{\rm CXB}$
for each annular region in table~\ref{tab:cxb_fluc},
where $\sigma$ is the standard
deviation of the CXB intensity $I_{\rm CXB}$.

\subsection{Non X-ray Background}
\label{sec:nxb}
The non X-ray background (NXB) spectra were estimated from the 
database of Suzaku night-earth observations using the procedure of \citet{tawa08}.  
We accumulated the data for the same detector area and 
the same distribution of COR2 as the A2142 observations,
 using an  FTOOL {\it xisnxbgen}. 
The night-earth data cover 150 days before and after the period of A2142 observations.  
To increase the signal-to-noise ratio 
by keeping the NXB count rate low, we selected durations in which
COR2 is $> 8$ GV.  
The systematic error due to the NXB uncertainty was estimated by varying the NXB intensity by $\pm 3\%$ as in
\citet{tawa08}.

\begin{figure*}[ht]
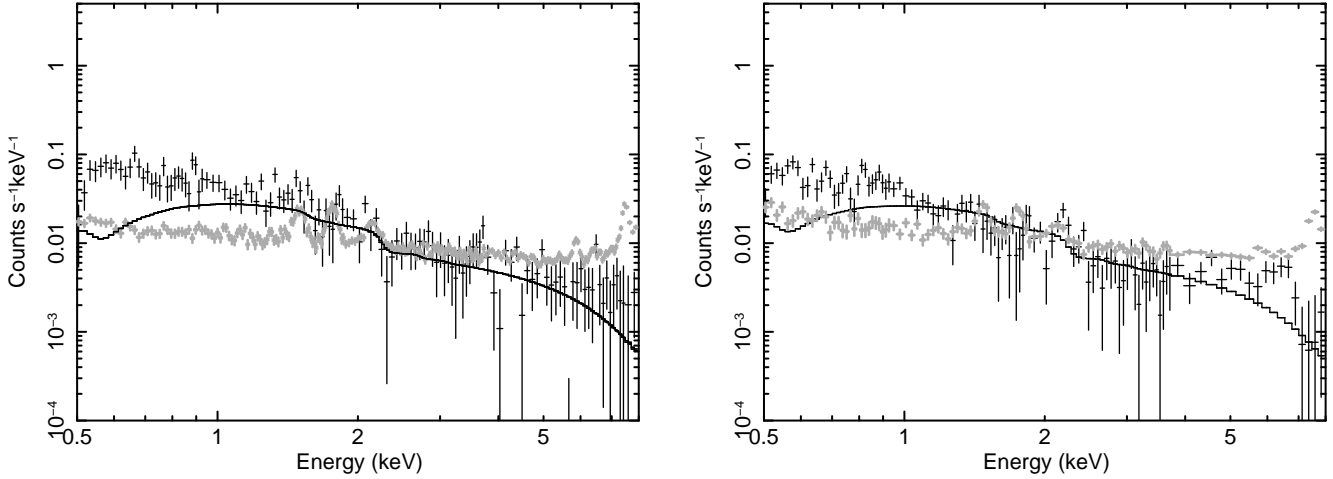

\begin{tabular}{cc}
\begin{minipage}{0.5\hsize}
\begin{center}
 \includegraphics[angle=-90,scale=0.35]{BGD-check-a2142off3-nops-bi.ps}
\label{fig:}
\end{center}
\end{minipage}
\begin{minipage}{0.5\hsize}
\begin{center}
 \includegraphics[angle=-90,scale=0.35]{BGD-check-tcrb-nops-bi.ps}
\end{center}
\end{minipage}
\end{tabular}
\caption{NXB-subtracted XIS BI spectra (Black cross) of Offset3 (left) and TCrB(right), 
plotted with estimated CXB (black line) and NXB (gray cross) spectrum described in Sec~\ref{sec:cxb}, 
Sec~\ref{sec:nxb}  respectively. }
\label{fig:bgd}
\end{figure*}

\subsection{Galactic Components}\label{sec:GAL}
To estimate the Galactic emission, we examined the spectra from two Suzaku observations: 
Offset3, which showed negligible ICM contribution, and TCrB at 1$^{\circ}$ south of A2142.  
We employed ancillary response files (ARFs) for a spatially uniform source
filling the FOV\@.  
A power-law is used to model the CXB in both spectra.  
The Galactic emission is represented
as a two-temperature model consisting of an unabsorbed 
$\sim$0.1 keV plasma 
(LHB; representing the local hot bubble and the solar wind charge exchange) and 
an absorbed $\sim$0.3 keV plasma (MWH; representing the Milky Way halo): 
${\it apec}_1+{\it wabs} \times ({\it apec}_2 + {\it powerlaw})$.  
The redshift and abundance of both the {\it apec} components were fixed at 0 and unity, respectively.

The best-fit parameters for the Offset3 and TCrB regions are summarized in table \ref{tab:bgd}. 
The temperatures of the LHB and the MWH are  
$0.09\pm0.02$ keV and $0.28\pm0.04$ keV, respectively, for Offset3, and 
$0.09\pm0.02$ keV and $0.29\pm0.03$ keV, respectively, for TCrB background.  
The temperatures and intensities are consistent with the typical Galactic emission.  
The observed fluxes in the two regions differ by 8\%, higher in Offset3.
We assume that the difference of the two fluxes 
represents the typical uncertainty range and the systematic error of
the Galactic emission.

\begin{table*}[ht]
  \caption{
  Best-fit values of background spectra fitting
  }
\begin{center}
\begin{tabular}{lcccccccccccc}\hline
\multicolumn{7}{c}{
A2142 OFFSET3
}\\ 
\hline 
 &  Unabsorbed (keV)	&$norm_1^{\ast}$  	& $S^{\dagger} _{1 \rm[0.4-10.0 keV]}$ &Absorbed (keV)&$norm_2^{\ast}$  & $S^{\dagger}_{ 2\rm[0.4-10.0 keV]}$ \\ \hline
nominal   & $  0.090 ^{+0.020}_{-0.018} $ & $ 4.90 ^{+4.76}_{-2.71}$ & 0.59$\pm$0.02
&$ 0.275 ^{+0.041}_{-0.038} $ & $ 0.51 ^{+0.20}_{-0.12} $ & 0.68$\pm$0.02 \\
CONTAMI+10\%  & $  0.090 ^{+0.013}_{-0.019} $ & $ 6.13
^{+10.41}_{-2.98}$ &  0.68$\pm$0.02
&$ 0.279 ^{+0.035}_{-0.039} $ & $ 0.55^{+0.22}_{-0.12} $ &0.74$\pm$0.02 \\
CONTAMI-10\%  & $  0.095 ^{+0.023}_{-0.021} $ & $ 3.27
^{+2.23}_{-1.67}$ & 0.53$\pm$0.02
&$ 0.281 ^{+0.040}_{-0.041} $ & $ 0.47
^{+0.18}_{-0.11} $ & 0.63$\pm$0.02\\

NXB+3\%+MAXCXB& $  0.092 ^{+0.022}_{-0.023} $ & $ 4.01
^{+6.15}_{-2.19}$ &0.63$\pm$0.02
&$ 0.266 ^{+0.040}_{-0.038} $ & $ 0.53
^{+0.23}_{-0.13} $ & 0.72$\pm$0.02\\
NXB-3\%+MINCXB & $  0.094 ^{+0.013}_{-0.019} $ & $ 4.17
^{+3.20}_{-2.01}$ & 0.59$\pm$0.02
&$ 0.286 ^{+0.038}_{-0.041} $ & $ 0.47
^{+0.18}_{-0.11} $ & 0.66$\pm$0.02\\ \hline \hline
\multicolumn{7}{c}{
TCrB
}\\ \hline
& \multicolumn{3}{c}{Unabsorbed plasma} & \multicolumn{3}{c}{Absorbed plasma} \\
 &  kT (keV)	&$norm_1^{\ast}$  	& $S^{\dagger} _{1 \rm[0.4-10.0 keV]}$ &kT (keV)&$norm_2^{\ast}$  & $S^{\dagger}_{ 2\rm[0.4-10.0 keV]}$ \\ \hline
  nominal  & $  0.089 ^{+0.010}_{-0.016} $ & $ 3.85 ^{+5.14}_{-1.32}$
& 0.41$\pm$0.01
&$ 0.292 ^{+0.032}_{-0.026} $ & $ 0.52 ^{+0.09}_{-0.10} $ &0.72$\pm$0.02 \\
COMTAMI+10\% & $  0.088 ^{+0.008}_{-0.016} $ & $ 4.48
^{+6.86}_{-1.34}$ & 0.35$\pm$0.01
&$ 0.291 ^{+0.031}_{-0.027} $ & $ 0.51
^{+0.09}_{-0.09} $ &  0.71$\pm$0.02\\
CONTAMI-10\% & $  0.087 ^{+0.019}_{-0.013} $ & $ 3.36
^{+3.82}_{-1.62}$ & 0.42$\pm$0.01
&$ 0.285 ^{+0.050}_{-0.023} $ & $ 0.45
^{+0.07}_{-0.12} $ & 0.72$\pm$0.02\\
NXB+3\%+MAXCXB & $  0.086 ^{+0.014}_{-0.015} $ & $ 4.24
^{+6.47}_{-1.66}$ & 0.44$\pm$0.01
&$ 0.270 ^{+0.042}_{-0.020} $ & $ 0.53
^{+0.10}_{-0.13} $ &  0.70$\pm$0.02\\
NXB-3\%+MINCXB  & $  0.089 ^{+0.009}_{-0.016} $ & $ 3.95
^{+5.43}_{-1.27}$ & 0.34$\pm$0.01
&$ 0.293 ^{+0.030}_{-0.025} $ & $ 0.52
^{+0.09}_{-0.09} $ &  0.60$\pm$0.02\\
 
 \hline \hline
\multicolumn{7}{l}{\footnotesize
*:\parbox[t]{160mm}{
Normalization of the apec component scaled with a factor 1/400$\pi$ 
assumed in the uniform-sky ARF calculation (circle radius $r$=\timeform{20'}).
Norm=$\rm \frac{1}{400\pi}$$\int n_{\rm e}n_{\rm H}{\rm dV}/(4\pi(1+z^2)D_{A}^2)\times 10 ^{-20} ~\rm cm^{-5}~arcmin^{-2}$, where $D_A$ is the angular diameter distance to the source.}}\\
\multicolumn{7}{l}{\footnotesize
$\dagger$: $10^{-6}~\rm photons~cm^{-2} s^{-1} arcmin^{-2}$. Energy band is 0.4 -10.0 keV.}\\
\label{tab:bgd}
\end{tabular}
\end{center}
\end{table*}

\subsection{Background Fraction in Each Region}
Table~\ref{tab:cxb_fluc} summarize the information in each annular
region we analyzed.  The columns indicate: the annular
boundaries; $\Omega_{\rm e}$, the solid angle
of observed areas; {\it Coverage}, 
the coverage fraction of each annulus, which is the
ratio of $\Omega_{\rm e}$ to the total solid angle of
the annulus; 
{\it SOURCE\_RATIO\_REG}, the fraction of the simulated cluster photons 
that fall in the region compared with the total photons generated in the
entire simulated cluster; $\sigma/I_{\rm CXB}$, the CXB
fluctuation due to unresolved point sources;
OBS, the observed counts; the estimated counts for the three background
components, i.e., NXB, CXB, and the Galactic emission; and the fraction of background photons
given by $f_{\rm BGD}\equiv$ (NXB+CXB+Galactic)/OBS\@.

The NXB counts are calculated from the night earth data.  
We simulated the spectra of the Galactic and CXB components, using {\it xissim}
\citep{ishisaki07} with the flux and spectral parameters given in the row
of ``A2142 OFFSET3 nominal'' in table \ref{tab:bgd}, assuming a uniform
surface brightness that fills the XIS FOV.  
We plot the NXB and CXB BI spectra compared with the observed spectra in the background regions in figure \ref{fig:bgd}.  
These simulated spectra gave the counts shown in table~\ref{tab:cxb_fluc}.

\begin{table*}[ht]
\begin{center}
\caption{Estimation of CXB fluctuation and Backgrand Count}
\label{tab:cxb_fluc}
\footnotesize
\begin{tabular}{cccccccccc}
\hline
Region   & $\it\Omega_e^{\ast}$ &$Coverage^{\dagger}$ &$SOURCE\_$ & $\sigma/I_{CXB}^\S$ &\multicolumn{5}{c}{FI Count 
(0.5-10 keV)}\\ 
		&	(arcmin$^2$)&	 (\%)	&$Ratio\_ Reg^{\ddagger}$ &(\%)	&OBS ($\times10^{2}$) &NXB  ($\times10^{2}$) & CXB ($\times10^{2}$) &Galactic ($\times10^{2}$)  &$f_{BGD}$(\%)	  \\ \hline \hline
\timeform{0'-2.5'} & 21.0 & 100.0 &  27.0 & 24.1  & 1701.8 $ \pm $ 4.1 & 4.1 $ \pm $ 0.1 &3.9  $ \pm $ 0.9 & 1.5 $ \pm $ 0.1 &0.6 $ \pm $ 0.1  \\
\timeform{2.5'}-\timeform{5.0'} & 58.8 & 95.4 &  31.1 & 14.4  & 1131.2 $ \pm $ 3.4 & 11.5 $ \pm $ 0.3 &9.9  $ \pm $ 1.4 & 4.4 $ \pm $ 0.2 &2.3 $ \pm $ 0.4  \\
\timeform{5.0'}-\timeform{7.8'} & 96.3 & 93.6 &  16.0 & 11.3  & 326.8 $ \pm $ 1.8 & 10.8 $ \pm $ 0.3 &9.8  $ \pm $ 1.1 & 3.8 $ \pm $ 0.2 &7.5 $ \pm $ 1.4  \\
\timeform{7.8'}-\timeform{10.3'}& 126.6 & 86.2 &  7.7 & 9.8  & 122.0 $ \pm $ 1.1 & 9.2 $ \pm $ 0.3 &7.2  $ \pm $ 0.7 & 3.1 $ \pm $ 0.2 &16.1 $ \pm $ 3.1  \\
\timeform{10.3'}-\timeform{12.9'} & 128.9 & 67.6 &  3.4 & 9.7  & 36.9 $ \pm $ 0.8 & 6.1 $ \pm $ 0.2 &5.7  $ \pm $ 0.6 & 2.5 $ \pm $ 0.2 &38.7 $ \pm $ 8.2  \\
\timeform{12.9'}-\timeform{16.8'} & 101.8 & 42.5 &  1.1 & 10.3  & 31.8 $ \pm $ 0.6 & 7.8 $ \pm $ 0.2 &7.1  $ \pm $ 0.8 & 2.8 $ \pm $ 0.2 &55.7 $ \pm $ 11.2  \\
\timeform{16.8'}-\timeform{24.	6'}  & 242.9 & 22.2 &  0.8 & 7.2  & 21.7 $ \pm $ 0.5 & 8.5 $ \pm $ 0.3 &5.8  $ \pm $ 0.4 & 2.4 $ \pm $ 0.2 &77.2 $ \pm $ 15.6  \\
\timeform{24.6'}-\timeform{31.	1'}  & 191.6 & 15.8 &  0.2 & 8.1  & 38.9 $ \pm $ 0.6 & 17.6 $ \pm $ 0.5 &16.8  $ \pm $ 1.4 & 6.4 $ \pm $ 0.2 &104.9 $ \pm $ 17.1  \\
\timeform{31.1'}-\timeform{38.8'}  & 241.7 & 14.1 &  0.1 & 7.2  & 31.2 $ \pm $ 0.6 & 16.5 $ \pm $ 0.5 &11.9  $ \pm $ 0.9 & 5.2 $ \pm $ 0.2 &107.8 $ \pm $ 18.4  \\ \hline
Region   & $\it\Omega_e^{\ast}$ &$Coverage^{\dagger}$ &$Source\_$ & $\sigma/I_{CXB}^\S$ &\multicolumn{5}{c}{BI Count 
(0.5-8 keV)}\\ 
		&	(arcmin$^2$)&	(\%)	&$Ratio\_ Reg^{\ddagger}$ &(\%)	&OBS ($\times10^{2}$) &NXB  ($\times10^{2}$) & CXB ($\times10^{2}$) &Galactic ($\times10^{2}$)  &$f_{BGD}$(\%)	  \\ \hline \hline
\timeform{0'-2.5'} & 21.0 & 100.0 &  27.0 & 24.1  & 1042.6 $ \pm $ 3.2 & 3.0 $ \pm $ 0.1 & 1.9 $ \pm $ 0.5 & 1.2 $ \pm $ 0.1 &0.6 $ \pm $ 0.2  \\
\timeform{2.5'}-\timeform{5.0'} & 58.8 & 95.4 &  31.1 & 14.4  & 732.5 $ \pm $ 2.7 & 8.2 $ \pm $ 0.2 & 5.1 $ \pm $ 0.7 & 3.8 $ \pm $ 0.2 &2.3 $ \pm $ 0.5  \\
\timeform{5.0'}-\timeform{7.8'}& 95.8 & 92.5 &  15.5 & 11.5  & 245.4 $ \pm $ 1.6 & 9.8 $ \pm $ 0.3 & 6.1 $ \pm $ 0.7 & 4.6 $ \pm $ 0.2 &8.4 $ \pm $ 1.6  \\
\timeform{7.8'}-\timeform{10.3'} & 122.6 & 86.2 &  7.7 & 9.8  & 88.8 $ \pm $ 0.9 & 7.8 $ \pm $ 0.2 & 4.9 $ \pm $ 0.5 & 3.2 $ \pm $ 0.2 &17.8 $ \pm $ 3.7  \\
\timeform{10.3'}-\timeform{12.9'}& 125.9 & 64.3 &  3.0 & 9.4  & 30.5 $ \pm $ 0.7 & 4.8 $ \pm $ 0.1 & 3.3 $ \pm $ 0.3 & 2.2 $ \pm $ 0.2 &33.8 $ \pm $ 7.8  \\
\timeform{12.9'}-\timeform{16.8'}& 105.8 & 46.7 &  1.2 & 10.8  & 26.2 $ \pm $ 0.5 & 6.0 $ \pm $ 0.2 & 4.1 $ \pm $ 0.5 & 2.8 $ \pm $ 0.2 &49.2 $ \pm $ 10.7  \\
 \timeform{16.8'}-\timeform{24.6'}& 233.9 & 20.8 &  0.7 & 8.2  & 18.6 $ \pm $ 0.4 & 7.2 $ \pm $ 0.2 & 3.6 $ \pm $ 0.3 & 2.7 $ \pm $ 0.2 &72.8 $ \pm $ 15.6  \\
\timeform{24.6'}-\timeform{31.	1'} & 181.6 & 13.8 &  0.2 & 9.3  & 31.6 $ \pm $ 0.6 & 14.6 $ \pm $ 0.4 & 9.2 $ \pm $ 0.7 & 6.2 $ \pm $ 0.2 &94.9 $ \pm $ 16.8  \\
\timeform{31.1'}-\timeform{38.8'}& 225.7 & 11.1 &  0.1 & 7.6  & 24.6 $ \pm $ 0.5 & 12.7 $ \pm $ 0.4 & 6.8 $ \pm $ 0.5 & 5.0 $ \pm $ 0.2 &99.3 $ \pm $ 18.4  \\ \hline
\multicolumn{10}{l}{\footnotesize
$\ast$: Solid angle of each observed region.  
}\\
\multicolumn{10}{l}{\footnotesize
${\dagger}$: Fraction of each area to entire annulus.
}\\
\multicolumn{10}{l}{\footnotesize
$\ddagger$: Fraction of the simulated cluster photons which fall in the region compared with the total photons generated in the entire simulated cluster.
}\\
\multicolumn{10}{l}{\footnotesize
\S: CXB fluctuation due to unresolved point sources.
}\\
\end{tabular}
\end{center}
\end{table*}

\section{Stray Light}
Stray light is photons entering from outside of the FOV.
The Suzaku optics often show non-negligible stray light from
nearby bright X-ray sources \citep{serlemitsos07}. 
The extended telescope point spread function also makes
contamination of photons from a nearby sky.
We estimated the contamination flux
based on a ray tracing simulation {\it xissim}.
The result of our simulation is shown in table~\ref{tab:stray}.
For each observed annulus, we calculated
the fraction originated from each sky area (i.e., annulus).
We can see that the ``on-source'' fractions are naturally higher for the central pointing than for the other regions, 
since the on-source flux itself is higher.  
At the same time, 20--30\% of the detected photons are due to
the contamination, and they are mostly from adjacent regions.
Therefore, the results indicate that even though the flux contamination
is not negligible,
the origin is limited to nearby regions.  
We proceeded to the spectral analysis
without making correction for the stray light.

\begin{table*}[ht]
\begin{center}
\caption{Stray light contamination for the central, OFFSET1 and OFFSET2 pointings}
\label{tab:stray}
\begin{tabular}{ccccccc}
\hline
\multicolumn{7}{c}{CENTER } \\
\hline
Emission Weighted	& Detector&\timeform{0'-2.5'}  &\timeform{2.5'}-\timeform{5.0'}  &\timeform{5.0'}-\timeform{7.8'}
&\timeform{7.8'}-\timeform{10.3'} &\timeform{10.3'}-\timeform{12.9'}  \\ 
Radius (arcmin)	&/Sky 	&\\
\hline \hline
  $1.1_{-1.1}^{+1.5}$	&\timeform{0'-2.5'}  &81.5 \% & 17.9 \%  & 0.5 \% &  0.1 \% & 0.0 \% \\
  $3.4_{-0.9}^{+1.7}$	&\timeform{2.5'}-\timeform{5.0'}  &26.5 \% & 66.9 \%  & 6.5 \% &  0.2 \% & 0.0 \% \\
$5.9_{-0.7}^{+1.9} $	&\timeform{5.0'}-\timeform{7.8'}  & 5.3 \% & 33.8 \%  & 56.0 \% &  4.8 \% & 0.1 \% \\
 $8.4_{-0.7}^{+1.9}$	&\timeform{7.8'}-\timeform{10.3'}  &4.3 \% & 8.1 \%  & 31.4 \% &  51.9 \% & 4.3 \% \\
 $11.0_{-0.6}^{+1.9}$&\timeform{10.3'}-\timeform{12.9'}  &2.9 \% & 6.0 \%  & 6.8 \% &  29.1 \% & 55.2 \% \\ \hline
\multicolumn{7}{c}{OFFSET1 and OFFSET2 } \\
\hline
Emission Weighted & Detector &$<$\timeform{12.9'}& \timeform{12.9'}-\timeform{16.8'}  &\timeform{16.8'}-\timeform{24.6'}
& \timeform{24.6'}-\timeform{31.	1'}  &\timeform{31.1'}-\timeform{38.8'}\\ 
Radius (arcmin)	&/Sky	&\\ \hline\hline
---&$<$\timeform{12.9'} & 94.1\% &4.7 \%&0.9\%& 0.2\%& 0.0\% \\
$14.0_{-1.1}^{+2.8}$ &\timeform{12.9'}-\timeform{16.8'}  &	21.2\% &61.7\% &15.8\% &1.2\% &0.0 \% \\
$18.6_{-1.8}^{+5.9}$	&\timeform{16.8'}-\timeform{24.6'} & 1.8\% &17.8\% &67.4\% &13.0\% &0.0\%  \\
$26.3_{-1.8}^{+4.7}$&\timeform{24.6'}-\timeform{31.1'} & 1.3\%& 2.4\%& 32.6\%& 63.1\%& 0.5\% \\
$32.5_{-1.4}^{+6.3}$&\timeform{31.1'}-\timeform{38.8'} & 6.7\%& 0.0\% &20.0\% &13.3\% &60.0\%  \\ \hline
\end{tabular}
\end{center}
\end{table*}

\begin{figure*}[ht]
\begin{tabular}{ccc}
\begin{minipage}{0.333\hsize}
{(a) \timeform{0'-2.5'} }
\\[-0.5cm]
\begin{center}
 \includegraphics[angle=-90,scale=0.22]{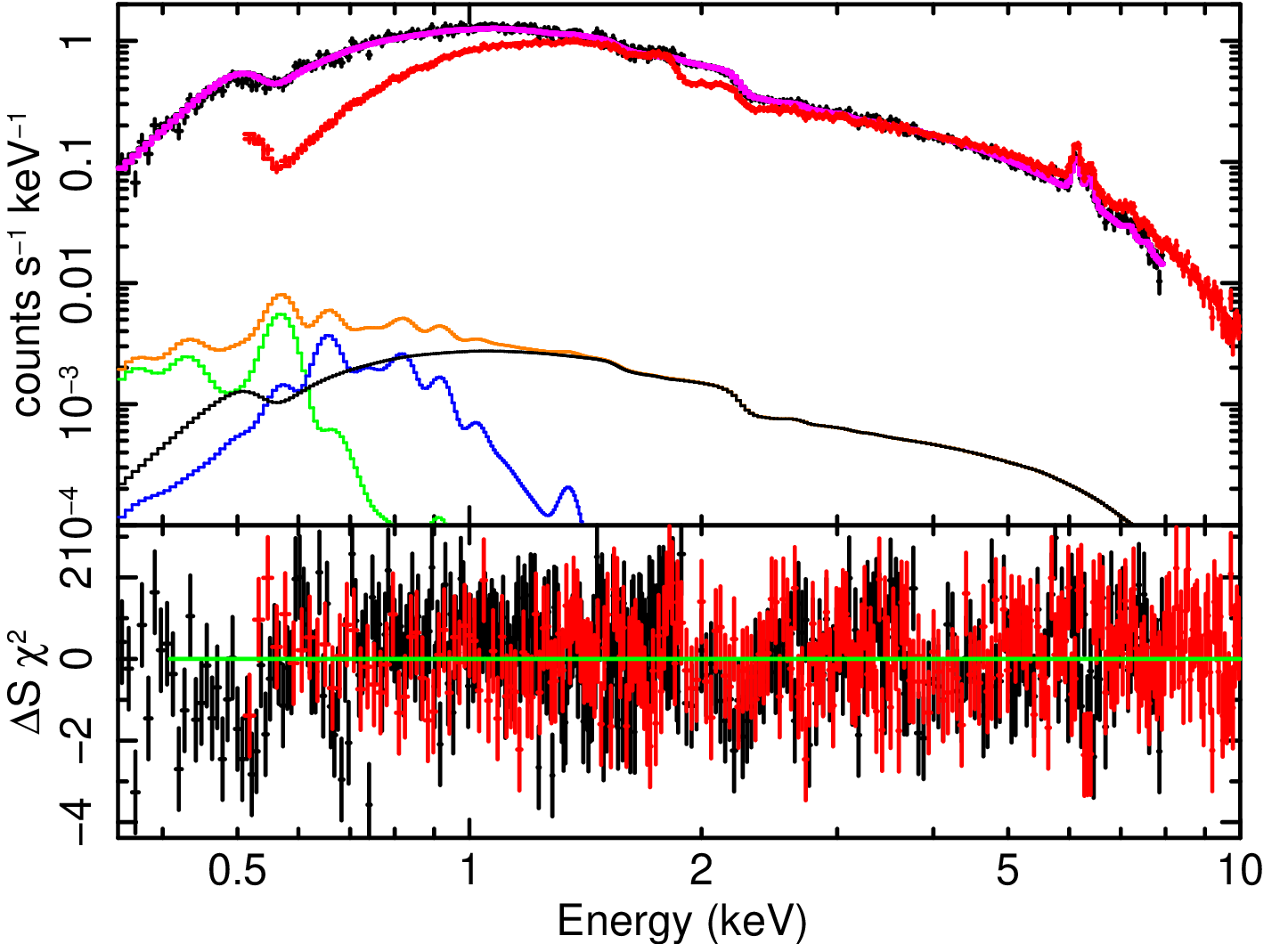}
\end{center}
\end{minipage}
\begin{minipage}{0.33\hsize}
{(b) \timeform{2.5'}-\timeform{5.0'} }
\\[-0.5cm]
\begin{center}
 \includegraphics[angle=-90,scale=0.22]{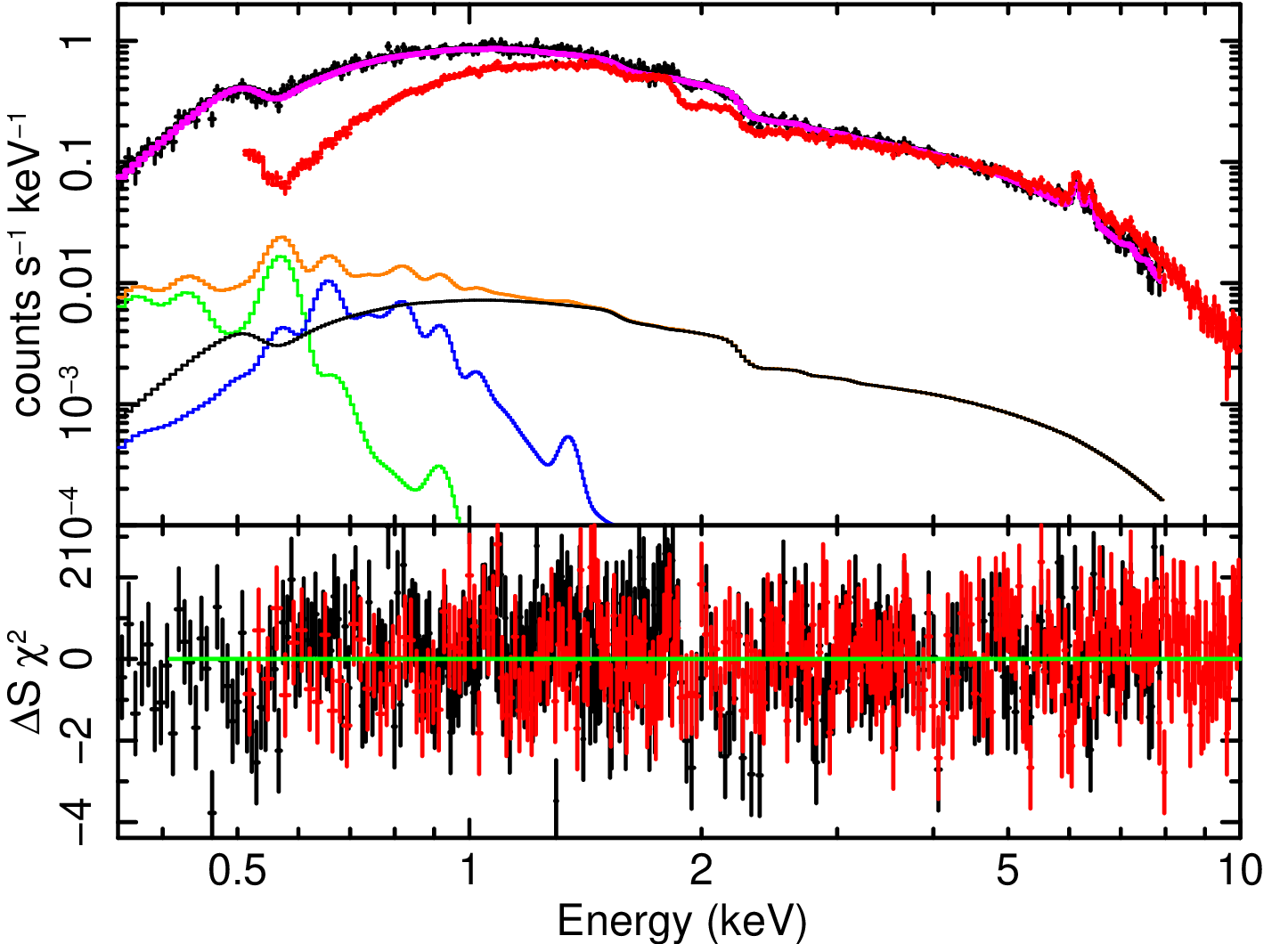}
\end{center}
\end{minipage}
\begin{minipage}{0.33\hsize}
{(c) \timeform{5.0'}-\timeform{7.8'} }
\\[-0.5cm]
\begin{center}
 \includegraphics[angle=-90,scale=0.22]{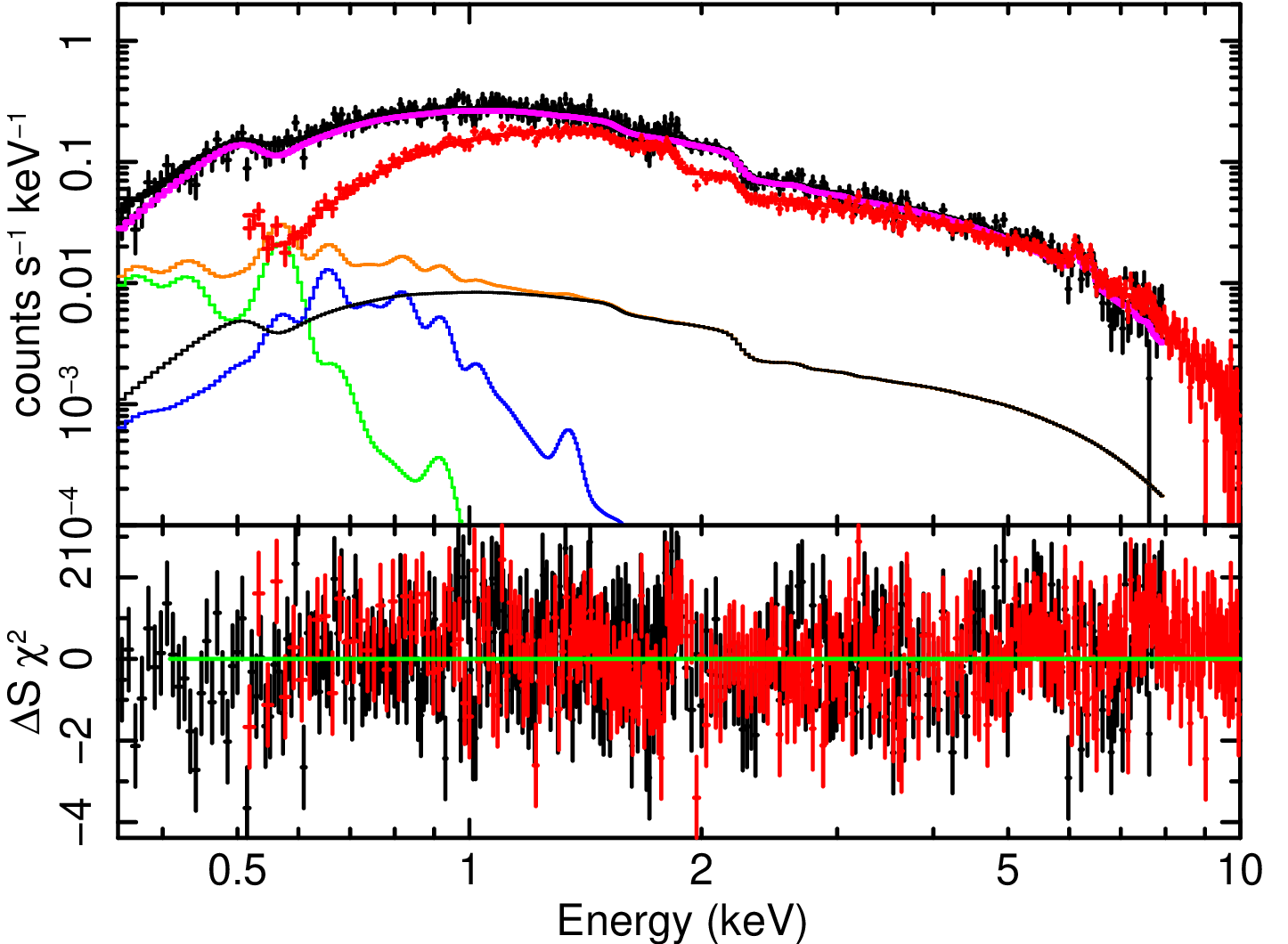}
\end{center}
\end{minipage}\\

\begin{minipage}{0.33\hsize}
{(d) \timeform{7.8'}-\timeform{10.3'} }
\\[-0.5cm]
\begin{center}
 \includegraphics[angle=-90,scale=0.22]{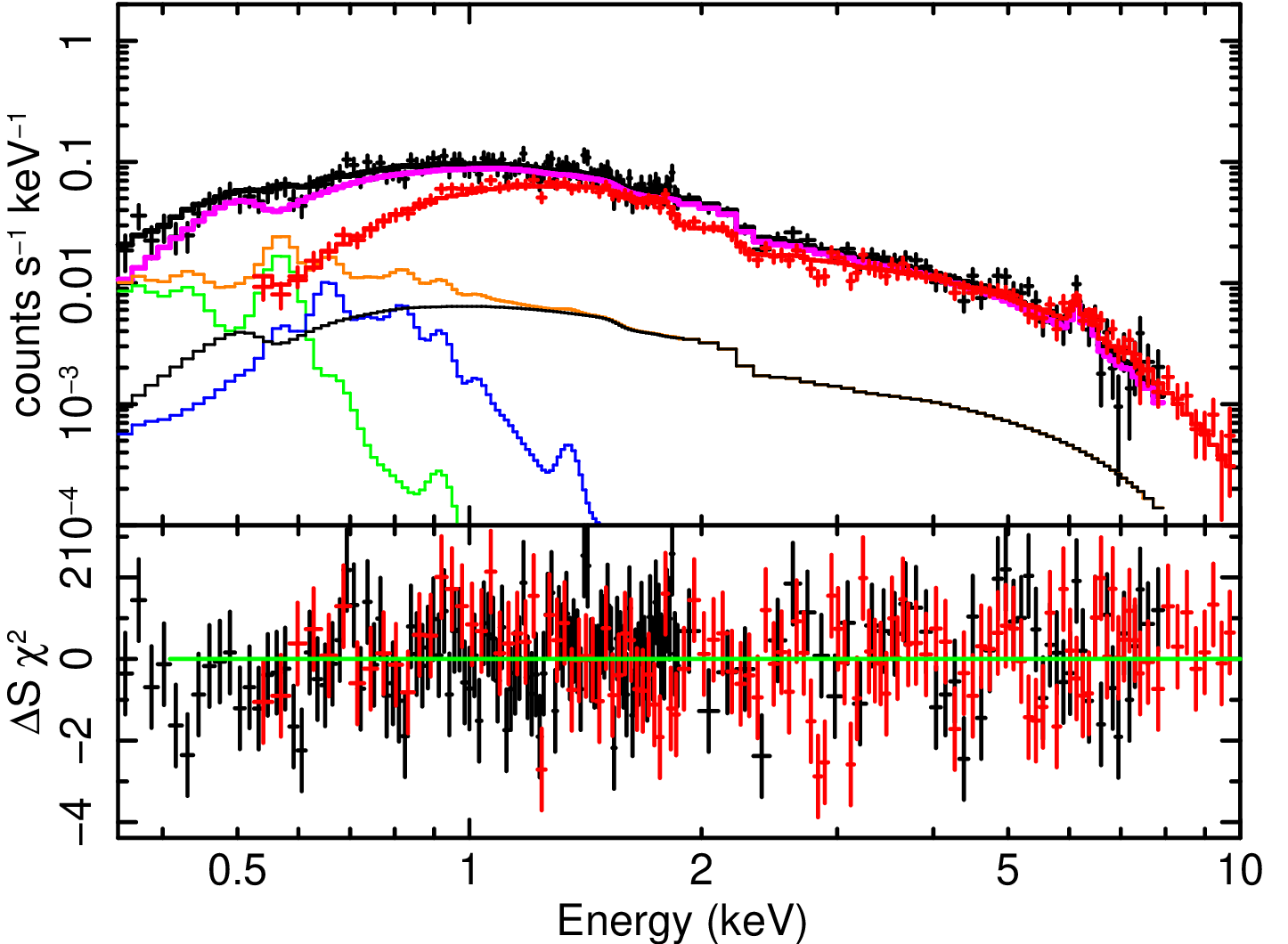}
\end{center}
\end{minipage}
\begin{minipage}{0.333\hsize}
{(e) \timeform{10.3'}-\timeform{12.9'} }
\\[-0.5cm]
\begin{center}
 \includegraphics[angle=-90,scale=0.22]{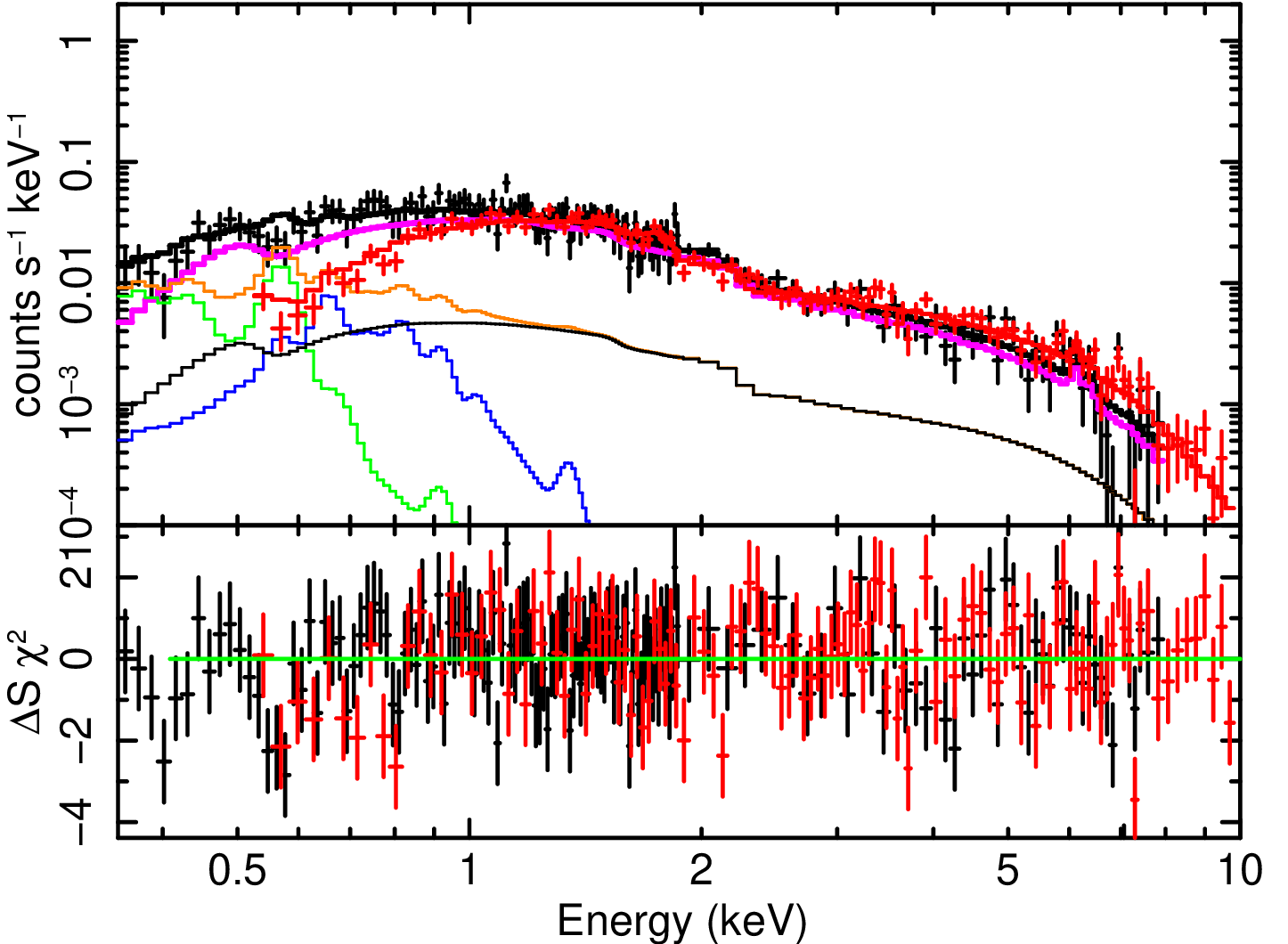}
\end{center}
\end{minipage}
\begin{minipage}{0.33\hsize}
{(f) \timeform{12.9'}-\timeform{16.8'} }
\\[-0.5cm]
\begin{center}
 \includegraphics[angle=-90,scale=0.22]{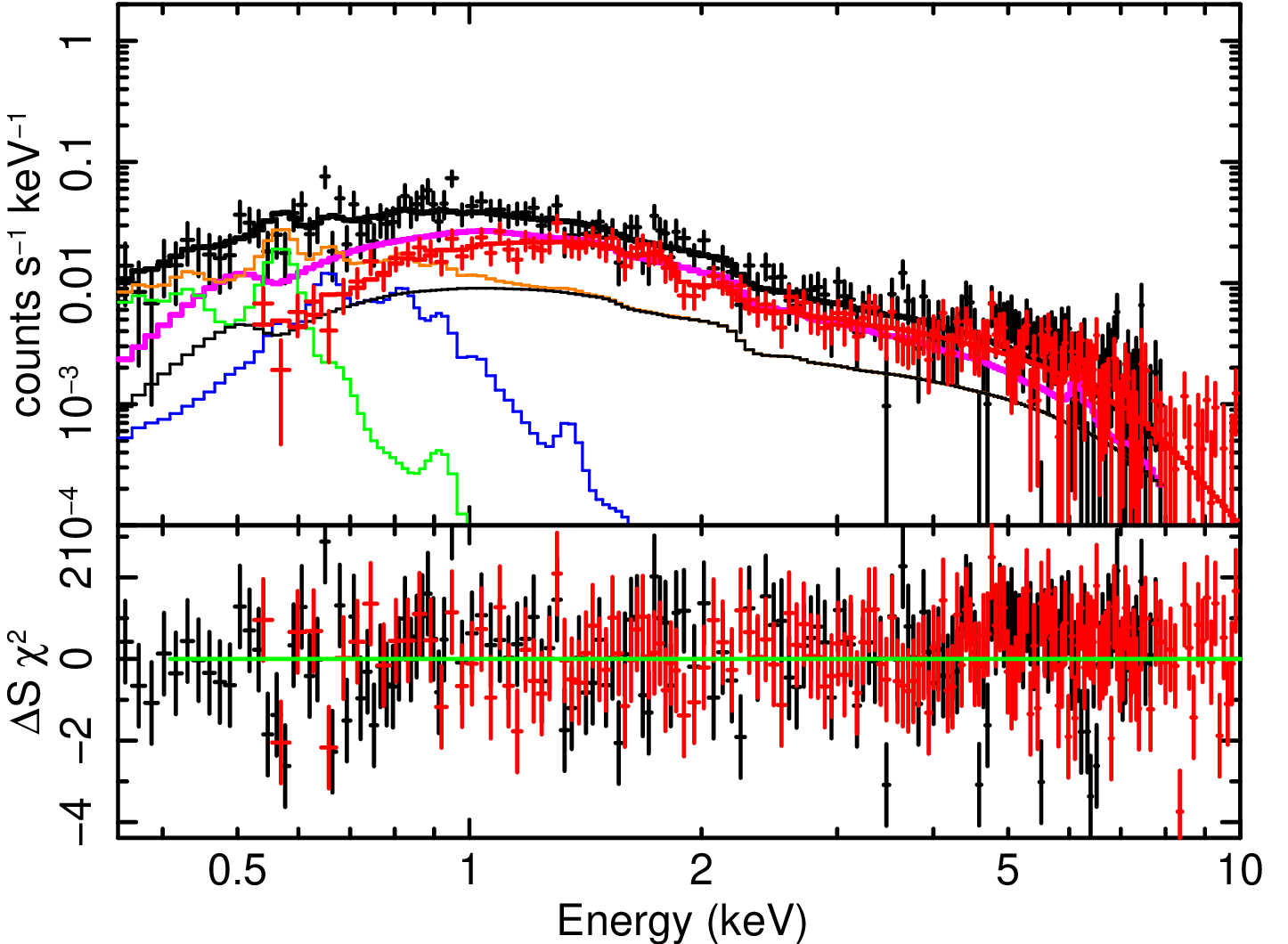}
\end{center}
\end{minipage}\\
\begin{minipage}{0.33\hsize}
{(f) \timeform{16.8'}-\timeform{24.6'}}
\\[-0.5cm]
\begin{center}
 \includegraphics[angle=-90,scale=0.22]{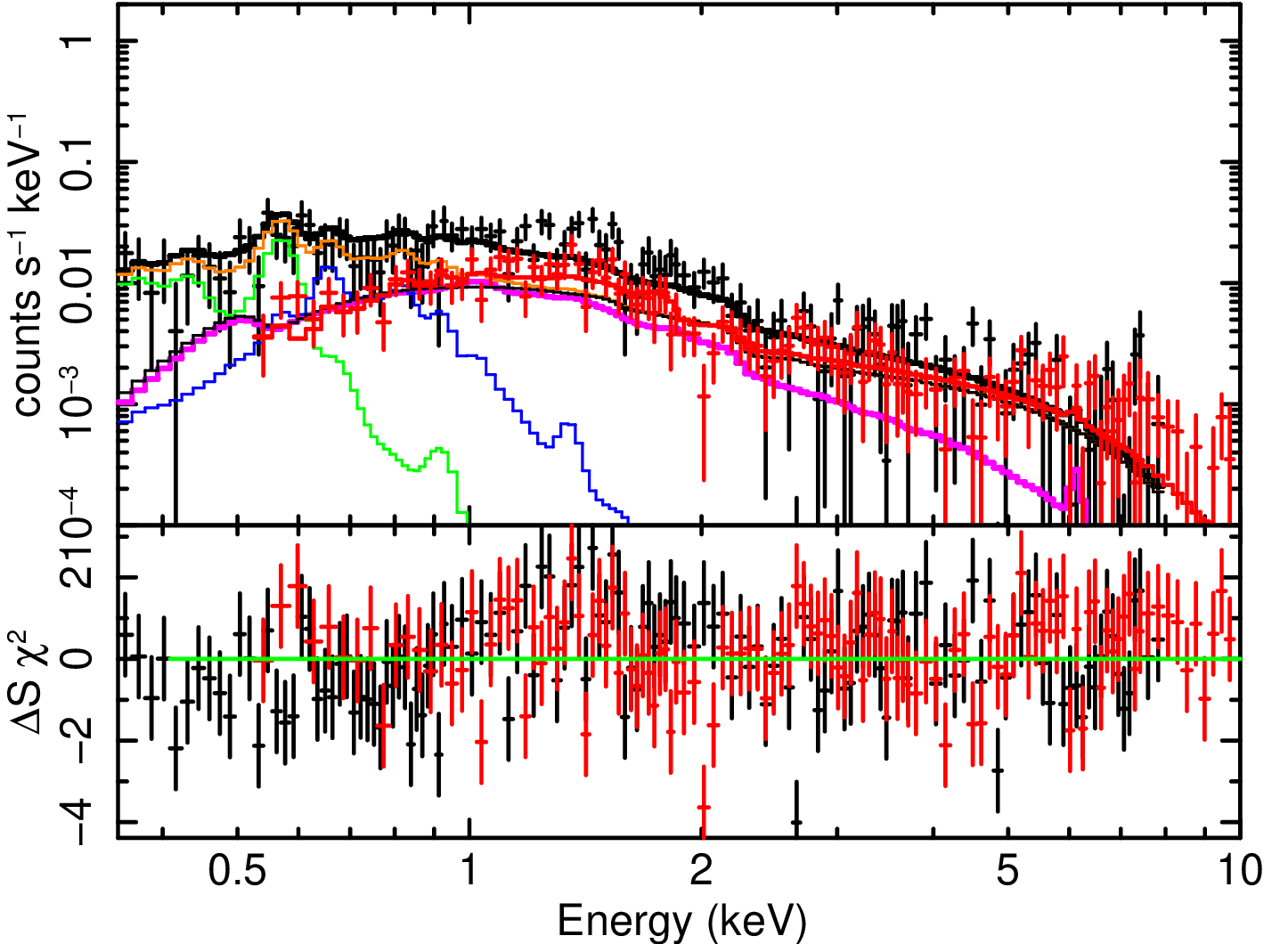}
\end{center}
\end{minipage}
\begin{minipage}{0.333\hsize}
{(g) \timeform{24.6'}-\timeform{31.1'} }
\\[-0.5cm]
\begin{center}
 \includegraphics[angle=-90,scale=0.22]{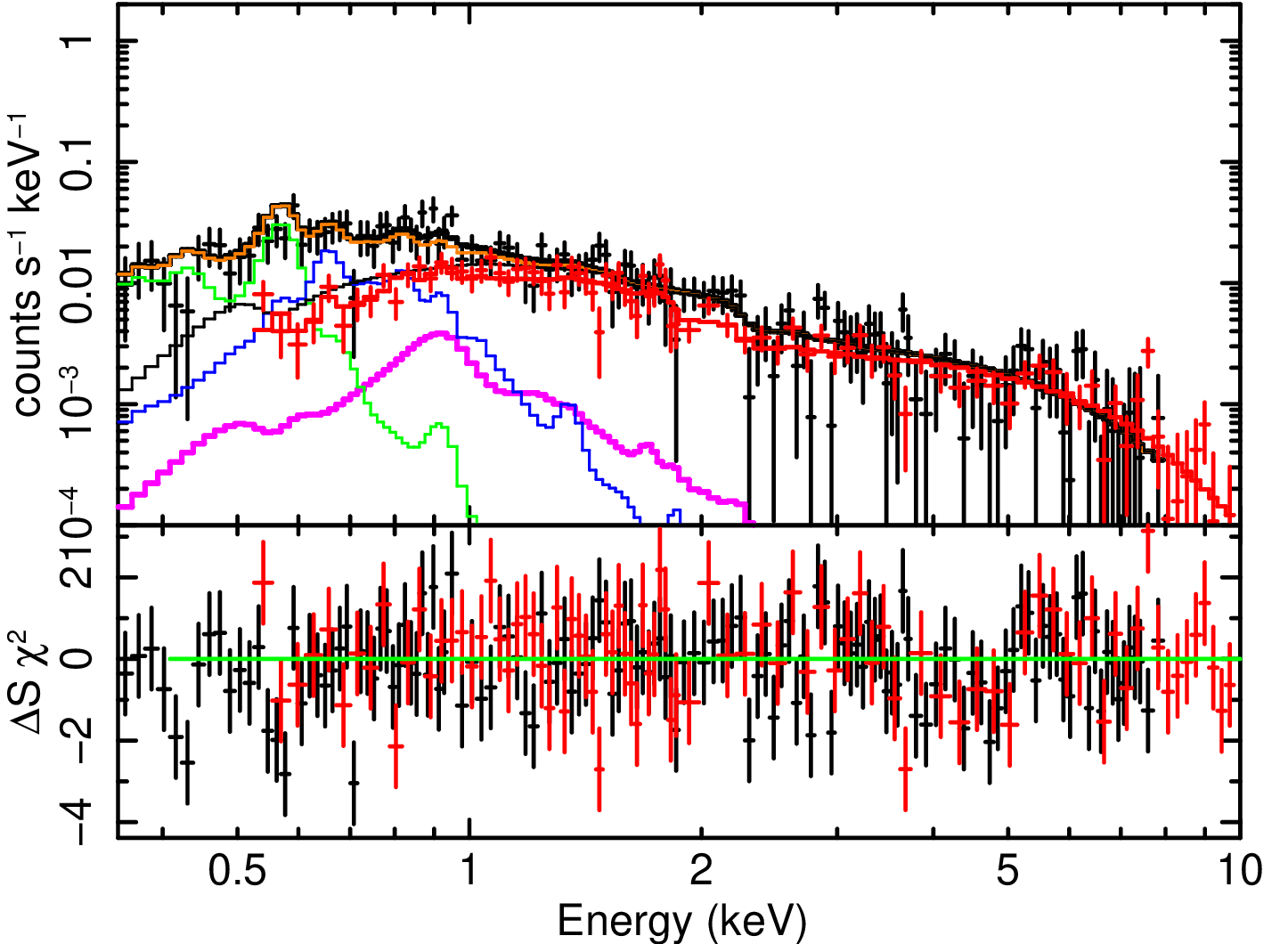}
\end{center}
\end{minipage}
\begin{minipage}{0.333\hsize}
{(h)\timeform{31.1'}-\timeform{38.8'} }
\\[-0.5cm]
\begin{center}
 \includegraphics[angle=-90,scale=0.22]{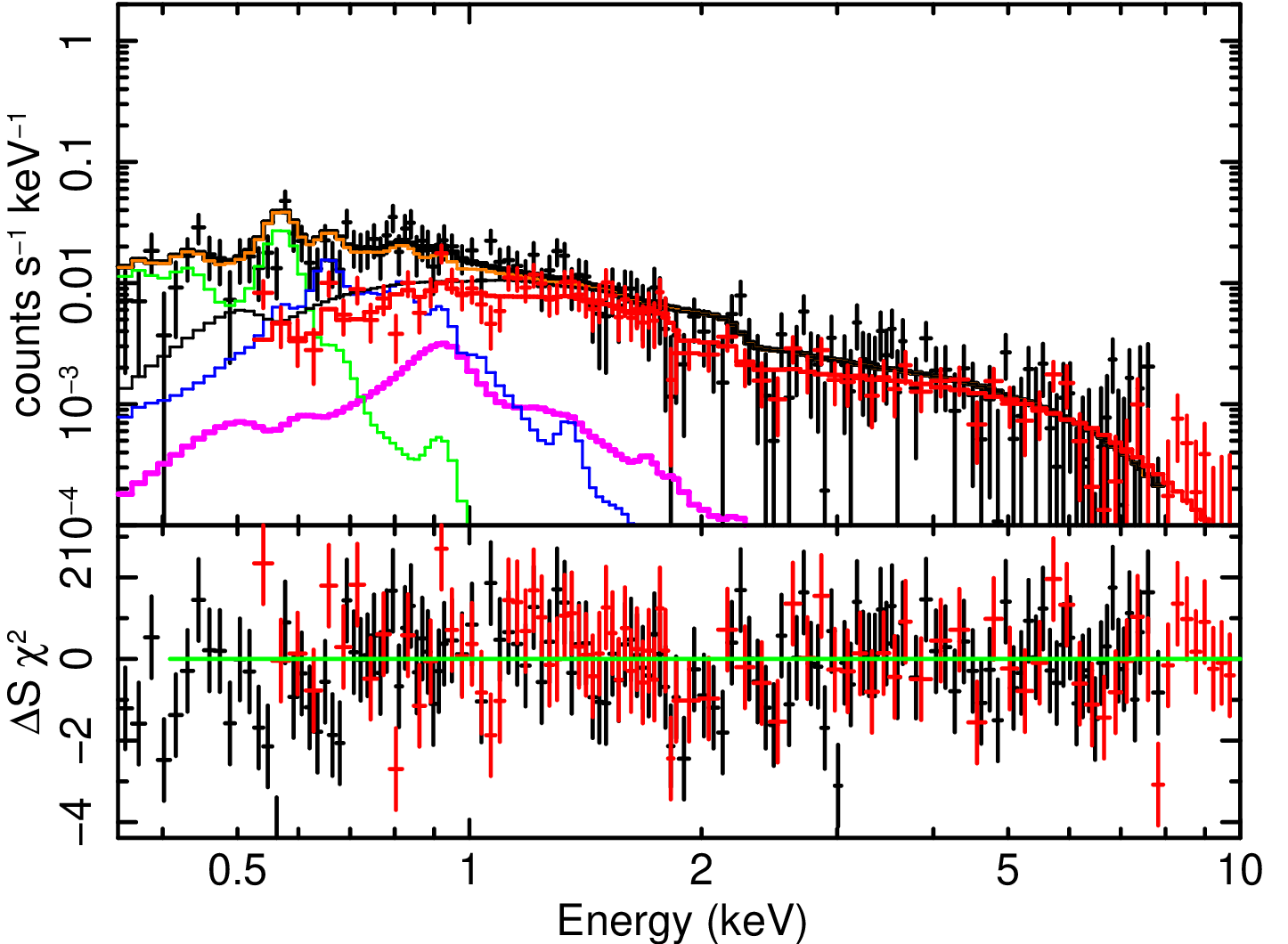}
\end{center}
\end{minipage}
\end{tabular}
\caption{ NXB subtracted spectrum for each region.  The XIS BI (Black)
  and FI (Red) spectra are fitted with ICM (${\it wabs + apec}$), added
  to CXB+the galactic components (LHB, MWH) ({\it apec+wabs(apec+powerlow)}). 
  The ICM components are shown in a magenta.
  The CXB components are shown in a black line, and the LHB and MWH emissions are indicated
  by green and blue lines, respectively.  }
\label{fig:fit}
\end{figure*}

\begin{table}[ht]
\tiny
\begin{center}
\caption{Best-fit parameters of the ICM with Offset3 background}
\label{tab:result_off3}
\begin{tabular}{cccccccc}
\hline
Region &k$T$ (keV) &$\it Z \rm (Z_{\odot})$&norm$^*$ & $S_{0.4-10 keV}^{\dagger}$ & BI/FI&$\chi^{2}$/d.o.f \\ \hline\hline
\timeform{0'-2.5'} & $  {8.16}^{+0.10}_{-0.10}  $  & $ {0.33}^{+0.02}_{-0.02} $ & $ {1.82}\pm 0.01 \times 10^{5}$ & $ {7.74}\pm 0.30 \times10^{4} $ & 0.92  &712 / 604\\
\timeform{2.5'}-\timeform{5.0'} & $  {9.02}^{+0.19}_{-0.19}  $  & $ {0.28}^{+0.02}_{-0.02} $ & $ {3.90}\pm 0.01 \times 10^{4} $ & $ {1.65}\pm0.05 \times10^{4} $ & 0.93  &691 / 604\\
\timeform{5.0'}-\timeform{7.8'} & $  {9.50}^{+0.38}_{-0.38}  $  & $ {0.24}^{+0.05}_{-0.05} $ & $ {7.50}\pm 0.01 \times10^{3}$ & $ {3.33}\pm 0.05\times10^{3} $ & 1.03  &681 / 604\\
\timeform{7.8'}-\timeform{10.3'} &$  {8.47}^{+0.65}_{-0.48}  $  & $ {0.31}^{+0.09}_{-0.09} $ & $ {2.09}\pm 0.07\times10^{3} $ & $ {9.59}\pm 0.33 \times10^{2}$ & 1.07  &309 / 270\\
\timeform{10.3'}-\timeform{12.9'} &$  {7.79}^{+0.81}_{-0.79}  $  & $ {0.19}^{+0.13}_{-0.12} $ & $ {8.30}\pm 0.32 \times10^{2}$ & $ {3.90}\pm 0.33 \times10^{2}$ & 1.19  &338 / 270\\
\timeform{12.9'}-\timeform{16.8'}& $  {6.14}^{+1.30}_{-0.95}  $  & 0.2 (fix)& $ {6.70}\pm 0.40 \times10^{2}$ & $ {2.67}\pm 0.22 \times10^{2}$ & 0.85  &381 / 358\\
\timeform{16.8'}-\timeform{24.6'}&$  {3.42}^{+2.14}_{-1.22}  $  & 0.2 (fix) & $ {1.35} \pm 0.22\times10^{2} $ & $ {5.18}\pm 0.22 \times10^{1}$ & 0.92  &272 / 217\\
\timeform{24.6'}-\timeform{31.1'} & $  {1.01}^{+0.44}_{-0.29}  $  & 0.2 (fix)& $ 2.00 \pm 0.11 \times 10^{1}$ & $ {6.40}\pm 0.20 $ & 0.94  &238 / 223\\
\timeform{31.1'}-\timeform{38.8'}& $  {1.08}^{+0.28}_{-0.37}  $  & 0.2 (fix)& $ 1.00 \pm 0.50 $ & $ {2.75}\pm 0.20$ & 0.88  &299 / 223\\
\hline
\multicolumn{7}{c}{CXBMAX} \\ \hline \hline
\timeform{0'-2.5'} & $  {8.16}^{+0.10}_{-0.10}  $  & $ {0.33}^{+0.02}_{-0.02} $ & $ {1.76}\pm 0.01\times 10^{5} $ & $ {7.52}\pm0.30\times10^{4} $ & 0.93  &712 / 604\\
\timeform{2.5'}-\timeform{5.0'} & $  {8.99}^{+0.19}_{-0.19}  $  & $ {0.28}^{+0.02}_{-0.02} $ & $ {3.80}\pm 0.05 \times10^{4} $ & $ {1.60}\pm 0.06\times10^{4} $ & 0.93  &690 / 604\\
\timeform{5.0'}-\timeform{7.8'} & $  {9.39}^{+0.38}_{-0.38}  $  & $ {0.23}^{+0.05}_{-0.05} $ & $ 7.37\pm 0.13 \times10^{3} $ & $ {3.23}\pm 0.05\times10^{3} $ & 1.03  &675 / 604\\
\timeform{7.8'}-\timeform{10.3'} &$  {8.36}^{+0.60}_{-0.48}  $  & $ {0.30}^{+0.09}_{-0.09} $ & $ {2.05}\pm 0.06 \times10^{3}$ & $ {9.35}\pm0.32 \times10^{2}$ & 1.07  &301 / 270\\
\timeform{10.3'}-\timeform{12.9'} &$  {7.65}^{+0.81}_{-0.79}  $  & $ {0.18}^{+0.12}_{-0.12} $ & $ {8.18}\pm 0.32 \times10^{2}$ & $ {3.82}\pm0.32 \times10^{2}$ & 1.18  &337 / 270\\
\timeform{12.9'}-\timeform{16.8'}& $  {6.26}^{+1.40}_{-1.00}  $  & 0.2 (fix)& $ {6.96}\pm 0.48 \times10^{2}$ & $ {2.75}\pm0.23 \times10^{2}$ & 0.84  &376 / 358\\
\timeform{16.8'}-\timeform{24.6'}& $  {3.96}^{+2.66}_{-1.27}  $  & 0.2 (fix)& $ {1.45}\pm 0.24 \times10^{2}$ & $ 5.64\pm0.29\times10^{1} $ & 0.90  &273 / 217\\
\timeform{24.6'}-\timeform{31.1'} & $  {1.06}^{+0.49}_{-0.31}  $  & 0.2 (fix)& $ {2.16}\pm1.21 \times10^{1}$ & $ {7.55}\pm0.40 $ & 0.89  &231 / 223\\
\timeform{31.1'}-\timeform{38.8'}& $  {1.02}^{+0.43}_{-0.47}  $  & 0.2 (fix)& $ {6.80}\pm 5.80 $ & $ {2.35}\pm0.03 $ & 0.82  &288 / 223\\ \hline

\multicolumn{7}{c}{CXBMIN} \\ \hline \hline
\timeform{0'-2.5'} & $  {8.17}^{+0.10}_{-0.10}  $  & $ {0.33}^{+0.02}_{-0.02} $ & $ {1.87}\pm0.01\times10^{5} $ & $ {7.98}\pm0.31\times10^{4} $ & 0.92  &713 / 604\\
\timeform{2.5'}-\timeform{5.0'} & $  {9.07}^{+0.19}_{-0.19}  $  & $ {0.28}^{+0.02}_{-0.02} $ & $ {4.04}\pm0.05\times10^{4} $ & $ {1.70}\pm0.06\times10^{4} $ & 0.93  &693 / 604\\
\timeform{5.0'}-\timeform{7.8'} & $  {9.65}^{+0.38}_{-0.38}  $  & $ {0.24}^{+0.05}_{-0.05} $ & $ {7.75}\pm0.13\times10^{3} $ & $ {3.44}\pm0.05\times10^{3} $ & 1.03  &687 / 604\\
\timeform{7.8'}-\timeform{10.3'} &$  {8.82}^{+0.70}_{-0.56}  $  & $ {0.31}^{+0.09}_{-0.09} $ & $ {2.19}\pm0.06\times10^{3} $ & $ {1.00}\pm0.03\times10^{3}$ & 1.07  &311 / 270\\
\timeform{10.3'}-\timeform{12.9'}&$  {8.34}^{+0.93}_{-0.78}  $  & $ {0.19}^{+0.13}_{-0.13} $ & $ {8.82}\pm0.30 \times10^{2}$ & $ {4.14}\pm0.34 \times10^{2}$ & 1.18  &339 / 270\\
\timeform{12.9'}-\timeform{16.8'}& $  {6.29}^{+1.32}_{-0.94}  $  & 0.2 (fix)& $ {7.69}\pm0.48 \times10^{2}$ & $ {2.33}\pm1.81 \times10^{2}$ & 1.19  &431 / 358\\
\timeform{16.8'}-\timeform{24.6'}& $  {3.38}^{+0.84}_{-0.62}  $  & 0.2 (fix)& $ {3.47}\pm0.25 \times10^{2}$ & $ {7.35}\pm 0.60\times10^{1} $ & 1.26  &255 / 217\\
\timeform{24.6'}-\timeform{31.1'} & $  {1.13}^{+1.40}_{-0.30}  $  & 0.2 (fix)& $ {1.97}\pm1.20\times10^{1} $ & $ 7.05\pm0.20 $ & 0.97  &242 / 223\\
\timeform{31.1'}-\timeform{38.8'}& $  {1.08}^{+0.43}_{-0.30}  $  & 0.2 (fix)& $ 7.20\pm5.20 $ & $ {2.55}\pm0.02 $ & 0.94  &304 / 223\\ \hline

\multicolumn{7}{c}{CONTAMI 10\% add} \\ \hline \hline
\timeform{0'-2.5'} & $  {7.85}^{+0.10}_{-0.10}  $  & $ {0.33}^{+0.02}_{-0.02} $ & $ {1.86}\pm0.01\times10^{5} $ & $ {7.86}\pm0.03\times10^{4} $ & 0.92  &734 / 604\\
\timeform{2.5'}-\timeform{5.0'} & $  {8.57}^{+0.19}_{-0.14}  $  & $ {0.28}^{+0.02}_{-0.02} $ & $ {3.94}\pm0.05\times10^{4} $ & $ {1.67}\pm0.06\times10^{4} $ & 0.93  &710 / 604\\
\timeform{5.0'}-\timeform{7.8'} & $  {9.03}^{+0.38}_{-0.37}  $  & $ {0.23}^{+0.04}_{-0.04} $ & $ {7.62}\pm0.13\times10^{3} $ & $ {3.73}\pm0.05\times10^{3} $ & 1.03  &685 / 604\\
\timeform{7.8'}-\timeform{10.3'} &$  {8.18}^{+0.50}_{-0.47}  $  & $ {0.30}^{+0.08}_{-0.08} $ & $ {2.15}\pm0.06\times10^{3} $ & $ {9.74}\pm0.29\times10^{2} $ & 1.06  &300 / 270\\
\timeform{10.3'}-\timeform{12.9'} &$  {7.57}^{+0.80}_{-0.77}  $  & $ {0.18}^{+0.12}_{-0.12} $ & $ {8.50}\pm0.03\times10^{2} $ & $ {3.97}\pm0.30\times10^{2} $ & 1.18  &325 / 270\\
\timeform{12.9'}-\timeform{16.8'}& $  {5.72}^{+1.06}_{-0.83}  $  & 0.2 (fix)& $ {6.96}\pm0.48 \times10^{2}$ & $ {2.73}\pm0.23 \times10^{2}$ & 0.84  &376 / 358\\
\timeform{16.8'}-\timeform{24.6'}& $  {3.22}^{+1.69}_{-0.94}  $  & 0.2 (fix)& $ {1.45}\pm0.24 \times10^{1}$ & $ {5.48}\pm0.27\times10^{1} $ & 0.91  &264 / 217\\
\timeform{24.6'}-\timeform{31.1'} & $  {1.08}^{+2.01}_{-0.46}  $  & 0.2 (fix)& $ {1.08}\pm0.90 \times10^{1}$ & $ {4.00}\pm1.00 \times10^{0}$ & 0.95  &255 / 223\\
\timeform{31.1'}-\timeform{38.8'}&  ----   & ----& -----  & ----- & 1:0.88  &319 / 223\\
\hline
\multicolumn{7}{c}{CONTAMI 10\% red} \\ \hline \hline
\timeform{0'-2.5'} & $  {8.27}^{+0.10}_{-0.10}  $  & $ {0.34}^{+0.02}_{-0.02} $ & $ {1.81}\pm0.01\times10^{5} $ & $ {7.71}\pm0.03\times10^{4} $ & 0.93  &764 / 604\\
\timeform{2.5'}-\timeform{5.0'} & $  {9.22}^{+0.19}_{-0.19}  $  & $ {0.28}^{+0.02}_{-0.02} $ & $ 3.89 \pm0.05\times10^{4}$ & $ {1.64}\pm0.05\times10^{4} $ & 0.94  &747 / 604\\
\timeform{5.0'}-\timeform{7.8'} & $  {9.81}^{+0.38}_{-0.38}  $  & $ {0.24}^{+0.05}_{-0.05} $ & $ {7.50}\pm0.13\times10^{3} $ & $ {3.31}\pm0.06 \times10^{3}$ & 1.04  &706 / 604\\
\timeform{7.8'}-\timeform{10.3'} &$  {8.80}^{+0.71}_{-0.54}  $  & $ {0.31}^{+0.09}_{-0.09} $ & $ {2.09}\pm0.06\times10^{3} $ & $ {9.54}\pm0.34 \times10^{2}$ & 1.07  &322 / 270\\
\timeform{10.3'}-\timeform{12.9'}&$  {8.39}^{+0.96}_{-0.81}  $  & $ {0.18}^{+0.14}_{-0.13} $ & $ {8.18}\pm0.32\times10^{2}$ & $ {3.86}\pm0.34 \times10^{2}$ & 1.19  &357 / 270\\
\timeform{12.9'}-\timeform{16.8'}& $  {6.60}^{+1.57}_{-1.08}  $  & 0.2 (fix)& $ {6.56}\pm0.40 \times10^{2}$ & $ {2.60}\pm0.20 \times10^{2}$ & 0.85  &392 / 358\\
\timeform{16.8'}-\timeform{24.6'}&  $  {4.15}^{+3.52}_{-1.38}  $  & 0.2 (fix) & $ {1.24}\pm0.24\times10^{2} $ & $ {4.89}\pm0.18\times10^{1} $ & 0.93  &286 / 217\\
\timeform{24.6'}-\timeform{31.1'} & $  {0.99}^{+0.37}_{-0.26}  $  & 0.2 (fix)& $ 1.91\pm1.10\times10^{1} $ & $ 7.00\pm0.20 $ & 0.94  &226 / 223\\
\timeform{31.1'}-\timeform{38.8'}& $  {1.06}^{+0.27}_{-0.34}  $  & 0.2 (fix)& $ 9.10\pm5.40 $ & $ 3.15\pm0.03 $ & 0.87  &283 / 223\\

\hline
\multicolumn{6}{l}{\tiny
*:Normalization of the apec component scaled with a factor
SOURCE\_RATIO\_REG$~{\it \Omega_e}$ from table ,} \\
\multicolumn{6}{l}{\tiny
Norm=$\rm \frac{SOURCE\_RATIO\_REG}{\it \Omega_e}\int n_{e}n_{H}
dV/(4\pi(1+z^2)D_{A}^2)\times 10^{-22} cm^{-5}~arcmin^{-2}$,}\\
\multicolumn{6}{l}{\tiny
where $D_A$ is the angular diameter distance to the source.}\\
\multicolumn{6}{l}{\tiny
$\dagger$: $\rm 10^{-8}photons cm^{-2} s^{-1}~arcmin^{-2}$. Energy band is 0.4 - 10.0 keV.}\\
\multicolumn{6}{l}{\tiny
Surface brightness of the apec component scaled with a factor SOURCE-RATIO-REG${\it \Omega_e}$ 
from
table~\ref{tab:cxb_fluc}.}
\end{tabular}
\end{center}
\end{table}

\begin{table}[ht]
\tiny
\begin{center}
\caption{Best-fit parameters of the ICM with TCrB background}
\label{tab:result_tcrb}
\begin{tabular}{cccccccc}
\hline
Region & k$T$(keV) &$\it Z\rm (Z_{\odot})$&norm$^*$ & $S_{0.4-10 keV}^{\dagger}$ &BI/FI& $\chi^{2}$/d.o.f
\\ \hline\hline
\timeform{0'-2.5'} & $  {8.16}^{+0.10}_{-0.10}  $  & $ {0.33}^{+0.02}_{-0.02} $ & $ 1.82 \pm 0.01\times10^{5} $ & $ {7.74}\pm 0.03\times10^{4} $ & 0.92  &711 / 604\\
\timeform{2.5'}-\timeform{5.0'} & $  {9.02}^{+0.19}_{-0.19}  $  & $ {0.28}^{+0.02}_{-0.02} $ & $ {3.89} \pm 0.05\times10^{4} $ & $ {1.65}\pm 0.06\times10^{4} $ & 0.93  &689 / 604\\
\timeform{5.0'}-\timeform{7.8'} & $  {9.47}^{+0.38}_{-0.38}  $  & $ {0.24}^{+0.05}_{-0.05} $ & $ {7.50} \pm 0.13\times10^{3} $ & $ {3.33}\pm 0.04\times10^{3}$ & 1.03  &674 / 604\\
\timeform{7.8'}-\timeform{10.3'} &$  {8.47}^{+0.66}_{-0.48}  $  & $ {0.31}^{+0.09}_{-0.09} $ & $ {2.12} \pm 0.07\times10^{3} $ & $ {9.63}\pm 0.29 \times10^{2}$ & 1.06  &298 / 270\\
\timeform{10.3'}-\timeform{12.9'}&$  {7.84}^{+0.82}_{-0.79}  $  & $ {0.19}^{+0.13}_{-0.12} $ & $ {8.39} \pm 0.32 \times10^{2}$ & $ {3.94}\pm 0.31 \times10^{2} $ &1.17  &323 / 270\\
\timeform{12.9'}-\timeform{16.8'}& $  {6.02}^{+1.21}_{-0.91}  $  & 0.2 (fix)& $ {6.88} \pm 0.48 \times10^{2}$ & $ {2.70}\pm 0.24 \times10^{2}$ & 0.84  &373 / 358\\
\timeform{16.8'}-\timeform{24.6'}& $  {3.62}^{+1.76}_{-1.14}  $  & 0.2 (fix) & $ {1.43} \pm 0.24\times10^{2}$ & $ {5.47}\pm 3.05 \times10^{1}$ & 0.89  &264 / 217\\
\timeform{24.6'}-\timeform{31.1'} & $  {1.07}^{+0.55}_{-0.31}  $  & 0.2 (fix)& $ {1.91} \pm 1.21\times10^{1} $ & $ {6.75}\pm 0.25 $ & 0.97  &228 / 223\\
\timeform{31.1'}-\timeform{38.8'}& $  {1.07}^{+0.35}_{-0.30}  $  & 0.2 (fix)& $ {9.11} \pm 6.07 $ & $ {3.15}\pm 0.25 $ & 0.87  &275 / 223\\
\hline
\multicolumn{6}{c}{CXBMAX} \\ \hline \hline
\timeform{0'-2.5'} & $  {8.16}^{+0.10}_{-0.10}  $  & $ {0.33}^{+0.02}_{-0.02} $ & $ {1.82} \pm 0.01\times10^{5} $ & $ {7.73}\pm 0.03\times10^{4} $ & 0.92  &711 / 604\\
\timeform{2.5'}-\timeform{5.0'} & $  {9.01}^{+0.19}_{-0.19}  $  & $ {0.28}^{+0.02}_{-0.02} $ & $ {3.89} \pm 0.05\times10^{4} $ & $ {1.65}\pm 0.06\times10^{4} $ & 0.93  &689 / 604\\
\timeform{5.0'}-\timeform{7.8'} & $  {9.46}^{+0.38}_{-0.38}  $  & $ {0.24}^{+0.05}_{-0.05} $ & $ {7.50} \pm 0.13 \times10^{3}$ & $ {3.31}\pm 0.05\times10^{3} $ & 1.03  &674 / 604\\
\timeform{7.8'}-\timeform{10.3'} &$  {8.46}^{+0.66}_{-0.48}  $  & $ {0.31}^{+0.09}_{-0.09} $ & $ {2.09} \pm 0.76\times10^{3} $ & $ {9.54}\pm 0.30 \times10^{2}$ & 1.06  &299 / 270\\
\timeform{10.3'}-\timeform{12.9'}& $  {7.81}^{+0.83}_{-0.80}  $  & $ {0.19}^{+0.13}_{-0.13} $ & $ {8.18} \pm 0.32 \times10^{2}$ & $ {3.86}\pm 0.31 \times10^{2}$ & 1.18  &324 / 270\\
\timeform{12.9'}-\timeform{16.8'}& $  {5.87}^{+1.20}_{-0.90}  $  & 0.2 (fix)& $ {6.56} \pm 0.40 \times10^{2}$ & $ {2.57}\pm0.22 \times10^{2}$ & 0.84  &373 / 358\\
\timeform{16.8'}-\timeform{24.6'}&$  {3.21}^{+1.93}_{-1.03}  $  & 0.2 (fix)& $ {1.28}\pm0.24 \times10^{1} $ & $ {4.81}\pm0.22\times10^{1} $ & 0.91  &267 / 217\\
\timeform{24.6'}-\timeform{31.1'} & $  {2.10}^{+0.56}_{-0.32}  $  & 0.2 (fix)& $ {7.00} \pm 1.55 $ & $ {2.55}\pm 0.15 $ & 0.91  &236 / 223\\
\timeform{31.1'}-\timeform{38.8'}&   --   & -- & -- & -- & 0.87& 287 / 223\\
\hline
\multicolumn{6}{c}{CXBMIN} \\ \hline \hline
\timeform{0'-2.5'} & $  {8.16}^{+0.10}_{-0.10}  $  & $ {0.33}^{+0.02}_{-0.02} $ & $ {1.82} \pm 0.01\times10^{5} $ & $ {7.74}\pm 0.30\times10^{4} $ & 0.92  &712 / 604\\
\timeform{2.5'}-\timeform{5.0'} & $  {9.03}^{+0.19}_{-0.19}  $  & $ {0.28}^{+0.02}_{-0.02} $ & $ {3.89} \pm 0.05\times10^{4} $ & $ {1.65}\pm 0.06\times10^{4} $ & 0.93  &689 / 604\\
\timeform{5.0'}-\timeform{7.8'} & $  {9.51}^{+0.38}_{-0.38}  $  & $ {0.24}^{+0.05}_{-0.05} $ & $ {7.62} \pm 0.13\times10^{3} $ & $ {3.34}\pm 0.51\times10^{3} $ & 1.03  &675 / 604\\
\timeform{7.8'}-\timeform{10.3'} &$  {8.53}^{+0.68}_{-0.47}  $  & $ {0.31}^{+0.09}_{-0.09} $ & $ {2.12} \pm 0.07\times10^{3} $ & $ {9.69}\pm 0.36\times10^{2}$ & 1.07  &299 / 270\\
\timeform{10.3'}-\timeform{12.9'}&$  {7.95}^{+0.84}_{-0.78}  $  & $ {0.19}^{+0.13}_{-0.12} $ & $ {8.50} \pm 0.32\times10^{2} $ & $ {3.98}\pm 0.32 \times10^{2}$ & 1.18  &324 / 270\\
\timeform{12.9'}-\timeform{16.8'}& $  {6.57}^{+1.52}_{-1.12}  $  & 0.2 (fix) & $ {7.28} \pm 0.48 \times10^{2}$ & $ {2.79}\pm 0.24 \times10^{2}$ & 0.84  &372 / 358\\
\timeform{16.8'}-\timeform{24.6'}& $ {3.96}^{+2.28}_{-1.18}  $  & 0.2 (fix) & $ {1.47}\pm0.22 \times10^{2}$ & $ {5.71}\pm 0.28\times10^{1}$ & 0.91  &268 / 217\\
\timeform{24.6'}-\timeform{31.1'} & $  {1.15}^{+1.56}_{-0.29}  $  & 0.2 (fix)& $ {2.23} \pm 0.14 \times10^{1}$ & $ {8.10}\pm 0.60 $ & 0.99  &222 / 223\\
\timeform{31.1'}-\timeform{38.8'}& $  {1.09}^{+0.42}_{-0.24}  $  & 0.2 (fix)& $ {9.57} \pm 6.30 $ & $ {3.35}\pm 0.15 $ & 0.92  &273 / 223\\
\hline
\multicolumn{6}{c}{CONTAMI 10\% ADD} \\ \hline \hline
\timeform{0'-2.5'} & $  {7.86}^{+0.10}_{-0.10}  $  & $ {0.33}^{+0.02}_{-0.02} $ & $ {1.86} \pm 0.01\times10^{5} $ & $ {7.86}\pm 0.32 \times10^{4}$ & 0.92  &734 / 604\\
\timeform{2.5'}-\timeform{5.0'} & $  {8.57}^{+0.20}_{-0.14}  $  & $ {0.28}^{+0.02}_{-0.02} $ & $ {3.94} \pm 0.05\times10^{4} $ & $ {1.67}\pm 0.06\times10^{4} $ & 0.93  &710 / 604\\
\timeform{5.0'}-\timeform{7.8'} & $  {9.05}^{+0.38}_{-0.37}  $  & $ {0.23}^{+0.04}_{-0.04} $ & $ {7.62} \pm 0.13\times10^{3} $ & $ {3.37}\pm 0.05\times10^{3} $ & 1.03  &684 / 604\\
\timeform{7.8'}-\timeform{10.3'} &$  {8.21}^{+0.52}_{-0.47}  $  & $ {0.30}^{+0.08}_{-0.08} $ & $ {2.15}\pm0.66\times10^{3} $ & $ {9.73}\pm0.25\times10^{2} $ & 1.06  &299 / 270\\
\timeform{10.3'}-\timeform{12.9'}&$  {7.64}^{+0.80}_{-0.79}  $  & $ {0.19}^{+0.12}_{-0.12} $ & $ {8.50}\pm0.32\times10^{2}  $ & $ {3.96}\pm0.32 \times10^{2} $ & 1.17  &322 / 270\\
\timeform{12.9'}-\timeform{16.8'}& $  {5.83}^{+1.10}_{-0.86}  $  & 0.2 (fix)& $ {6.88} \pm 0.48 \times10^{2}$ & $ {2.72}\pm 0.23 \times10^{2}$ & 0.84  &374 / 358\\
\timeform{16.8'}-\timeform{24.6'}& ${3.37}^{+1.93}_{-0.98}  $  & 0.2 (fix) & $ {1.42}\pm0.24\times10^{2} $ & $ {5.34}\pm0.25 \times10^{1}$ & 0.91  &267 / 217\\
\timeform{24.6'}-\timeform{31.1'} & $  {1.08}^{+0.62}_{-0.37}  $  & 0.2 (fix)& $ {1.53} \pm 1.15 \times10^{1} $ & $ {5.45}\pm 0.15 $ & 0.94  &227 / 223\\
\timeform{31.1'}-\timeform{38.8'}& $  {1.08}^{+0.41}_{-0.38}  $  & 0.2 (fix)& $ {7.24} \pm 6.30 $ & $ {2.45}\pm 0.15 $ & 0.88  &281 / 223\\

\hline
\multicolumn{6}{c}{CONTAMI 10\% RED} \\ \hline \hline
\timeform{0'-2.5'} & $  {8.27}^{+0.10}_{-0.10}  $  & $ {0.34}^{+0.02}_{-0.02} $ & $ {1.81} \pm 0.01\times10^{6} $ & $ {7.71}\pm 0.30\times10^{4} $ & 0.93  &760 / 604\\
\timeform{2.5'}-\timeform{5.0'} & $  {9.19}^{+0.19}_{-0.19}  $  & $ {0.28}^{+0.02}_{-0.02} $ & $ {3.89} \pm 0.05\times10^{5} $ & $ {1.65}\pm 0.06\times10^{4} $ & 0.94  &733 / 604\\
\timeform{5.0'}-\timeform{7.8'} & $  {9.67}^{+0.38}_{-0.38}  $  & $ {0.24}^{+0.05}_{-0.05} $ & $ {7.50} \pm 0.13\times10^{4} $ & $ {3.32}\pm 0.06\times10^{3} $ & 1.04  &683 / 604\\
\timeform{7.8'}-\timeform{10.3'} &$  {8.52}^{+0.68}_{-0.48}  $  & $ {0.30}^{+0.09}_{-0.09} $ & $ {2.12}\pm0.67\times10^{3}$ & $ {9.62}\pm0.32\times10^{2} $ & 1.07  &299 / 270\\
\timeform{10.3'}-\timeform{12.9'} & $  {7.92}^{+0.84}_{-0.79}  $  & $ {0.19}^{+0.13}_{-0.13} $ & $ {8.29}\pm0.32\times10^{2}  $ & $ {3.92}\pm0.33\times10^{2}  $ & 1.19  &317 / 270\\
\timeform{12.9'}-\timeform{16.8'}& $  {6.06}^{+1.23}_{-0.92}  $  & 0.2 (fix)& $ {6.80} \pm 0.40\times10^{2}  $ & $ {2.67}\pm 0.23 \times10^{2}  $ & 0.84  &372 / 358\\
\timeform{16.8'}-\timeform{24.6'}& $  {3.27}^{+1.71}_{-0.93}  $  & 0.2 (fix)& $ {1.45}\pm0.44\times10^{2}  $ & $ {5.45}\pm0.28 \times10^{1} $ & 0.90  &263 / 217\\
\timeform{24.6'}-\timeform{31.1'} & $  {0.97}^{+0.35}_{-0.25}  $  & 0.2 (fix)& $ {2.23} \pm 1.15\times10^{1} $ & $ {8.25}\pm 0.25 $ & 0.97  &228 / 223\\
\timeform{31.1'}-\timeform{38.8'}& $  {1.04}^{+0.26}_{-0.32}  $  & 0.2 (fix)& $ {1.10} \pm 0.58\times10^{1} $ & $ {3.85}\pm 0.25 $ & 0.87  &275 / 223\\
\hline
\multicolumn{6}{l}{\tiny
*:Normalization of the apec component scaled with a factor
SOURCE\_RATIO\_REG$~{\it \Omega_e}$ from table ,} \\
\multicolumn{6}{l}{\tiny
Norm=$\rm \frac{SOURCE\_RATIO\_REG}{\it \Omega_e}\int n_{e}n_{H}
dV/(4\pi(1+z^2)D_{A}^2)\times 10
^{-22} cm^{-5}arcmin^{-2}$,}\\
\multicolumn{6}{l}{\tiny
where $D_A$ is the angular diameter distance to the source.}\\
\multicolumn{6}{l}{\tiny
$\dagger$: $\rm 10^{-8}photons cm^{-2} s^{-1} arcmin^{-2}$. Energy band is 0.4 -
10.0 keV.}\\
\multicolumn{6}{l}{\tiny
Surface brightness of the apec component scaled with a factor
SOURCE-RATIO-REG${\it \Omega_e}$ from
table~\ref{tab:cxb_fluc}.}
\end{tabular}
\end{center}
\end{table}

\section{Spectral Analysis}\label{sec:specana}
\subsection{Spatial and Spectral Responses}\label{sec:resp}
We need to calculate the spatial and spectral responses for the analysis
of A2142 data.  The response functions for 
extended sources are complicated because 
they depend on the surface brightness distribution
of the source.  They need to be calculated for each annular region.
Monte Carlo simulator {\it xissim} incorporates the responses of the
X-ray telescope and XIS instrument.  
The ARF generator using this simulator is called {\it xissimarfgen} \citep{ishisaki07}.  
We used version 2008-04-05 of the simulator.  
The surface brightness distribution is one of the input parameters necessary to run {\it xissim} and {\it xissimarfgen}.
Because of the extended PSF, the local efficiency is related with  the relative flux among adjacent spatial elements.  
We used the $\beta$-model ($\beta=0.85, r_c =\timeform{4.5'}$) based on 
the ROSAT PSPC result as the input X-ray image~\citep{henry96}.

We created ARFs assuming that the input image does not vary with energy.
The effect of the contamination on the XIS IR/UV blocking filter is
included in the ARFs based on the calibration in November 2006.  
The normalization of the ARF is defined such that the flux given as a
result of the spectral fit is equal to the entire flux for a given spatial region.

\subsection{Spectral Fit}
We carried out spectral fitting to the pulse-height data of each
annular region separately.  The NXB component was subtracted before
the fit, and the fitting model included the LHB, MWH, CXB and ICM components.
The spectra from the BI and FI sensors were jointly fitted with the
same model by minimizing the total $\chi^{2}$ value.  
To increase the signal to noise ratio of A2142, 
we used energy ranges of 0.35--8 keV for BI and 0.5--10 keV for FI.  
The relative normalization between the two sensors was a free parameter
in this fit to compensate for the cross-calibration errors.
The photon index and normalization of the CXB, 
the temperatures and normalization of
the LHB and MWH were fixed at the values in table~\ref{tab:bgd}. 
Metal abundances of the LHB and MWH components were set to be unity. 
The Galactic absorption column density was fixed at $N_{\rm H}=4.2 \times 10^{20}\rm~cm^{-2}$~\citep{dickey90}.
We checked the influence of an uncertainty in $N_{\rm H}$ by using
another column density from Leiden/Argentine/Bonn (LAB) survey 
($N_{\rm H}=3.8 \times 10^{20}\rm~cm^{-2}$; \cite{kalberla05})
to confirm that the two spectrum fits do not show significant  difference.
The fits were carried out with XSPEC ver12.4.0ao.  
In the central regions, free parameters were the temperature, normalization, metal abundance of the ICM component.  
In the outer regions, we fixed the metal abundance of the ICM at 0.2, which is the lowest value observed in the outskirts of clusters
\citep{fujita08}.

Figure \ref{fig:fit} shows the results of the spectral fit for all the annular regions.  
The parameters and $\chi^2$ values are listed in tables \ref{tab:result_off3} and \ref{tab:result_tcrb}.  
We obtained fairly good fits for all the regions with reduced $\chi^{2}$ values less than 1.3.

\subsection{Temperature and Brightness Profiles}\label{sec:temp}
Figure~\ref{fig:parameters}(a) shows the radial profile of ICM
temperature, based on the result of the spectral fits.  The inner 5
annular regions have the width of about $\timeform{2'.5}$, and 20--40\% of the
detected flux comes from the adjacent sky regions.  The maximum
temperature within $\timeform{7'}$ (700 kpc) from the cluster center is $\sim 9$ keV, 
and it gradually decreases toward the outer region down to $\sim 4$ keV around the virial radius.  
There is a suggestion of weak emission with $kT \sim 1$ keV just outside of $r_{200}$.  
As shown in figure~\ref{fig:fit}(g) and (h), this emission (in magenta) has the
intensity much weaker than the other components throughout the energy range. 
If one takes into account the systematic error as shown in section~\ref{sec:err}, 
this component does not stay significant in the outermost region  in figure~\ref{fig:fit} (h) and in table \ref{tab:result_off3} and \ref{tab:result_tcrb}.  
Therefore, we treat the 1 keV emission in \timeform{31.1'}-\timeform{38.8'} to be an upper limit.

Also, we check projection effect on temperature profiles.
Based on \citet{fujita08}, we re-fit each annulus with the following method.
\begin{itemize}
\item Fitting the ICM spectrum of the outermost region
      (\timeform{24.6'}-\timeform{31.1'} ) with a single ICM component.
\item Fitting the next annulus, with a model of a combination of the
      ICM of this annulus and the overlapped emission from outer radius,
      i.e., the best fitting model of the outermost  region normalized to account for the spherical projection.
\item Proceeding to fitting of inner regions with a model of a combination of the
      ICM of this annulus and the overlapped emission from outer radius,
\end{itemize}
The results of this deprojection fitting do not show significant difference from the non-deprojection fitting.
In sec~\ref{sec:discussion}, we use the results of non-deprojection fitting method for discussion of ICM properties.

\subsection{Electron Density Profile}\label{sec:ne}
The electron density profile was calculated from the normalization
parameter of the Apec model, defined by
\begin{equation}
Norm=\frac{10^{-14}}{4\pi D_{A}^2(1+z)^2}\int n_e n_H dV
\end{equation}
with the unit of cm$^{-5}$, where $D_A$ (cm) is the angular diameter
distance to the source, $n_e$ and $n_{\rm H}$ (cm$^{-3}$) are the number
densities of electron and hydrogen, respectively. 
We note that the resultant normalization using an ARF generated by $\it xissimarfgen$ needs
the correction by a factor of SOURCE\_RATIO\_REG/$\Omega_{e}$.
See \cite{ishisaki07} Sec 5.3 for more information.
Each annular region,
projected in the sky, includes emission from different densities due
to integration along the line of sight.  We have de-convolved the
electron density assuming spherical symmetry by successively calculating
the value from the outermost regions. In this process, we assumed a
constant temperature in each annular region. The uncertainty 
of the density due to this assumption is a few \%.

\begin{figure}[ht]
\begin{tabular}{c}
\begin{minipage}{1.0\hsize}
(a) Temperature
\\[-.8cm]
\begin{center}
 \includegraphics[angle=-90,scale=0.4]{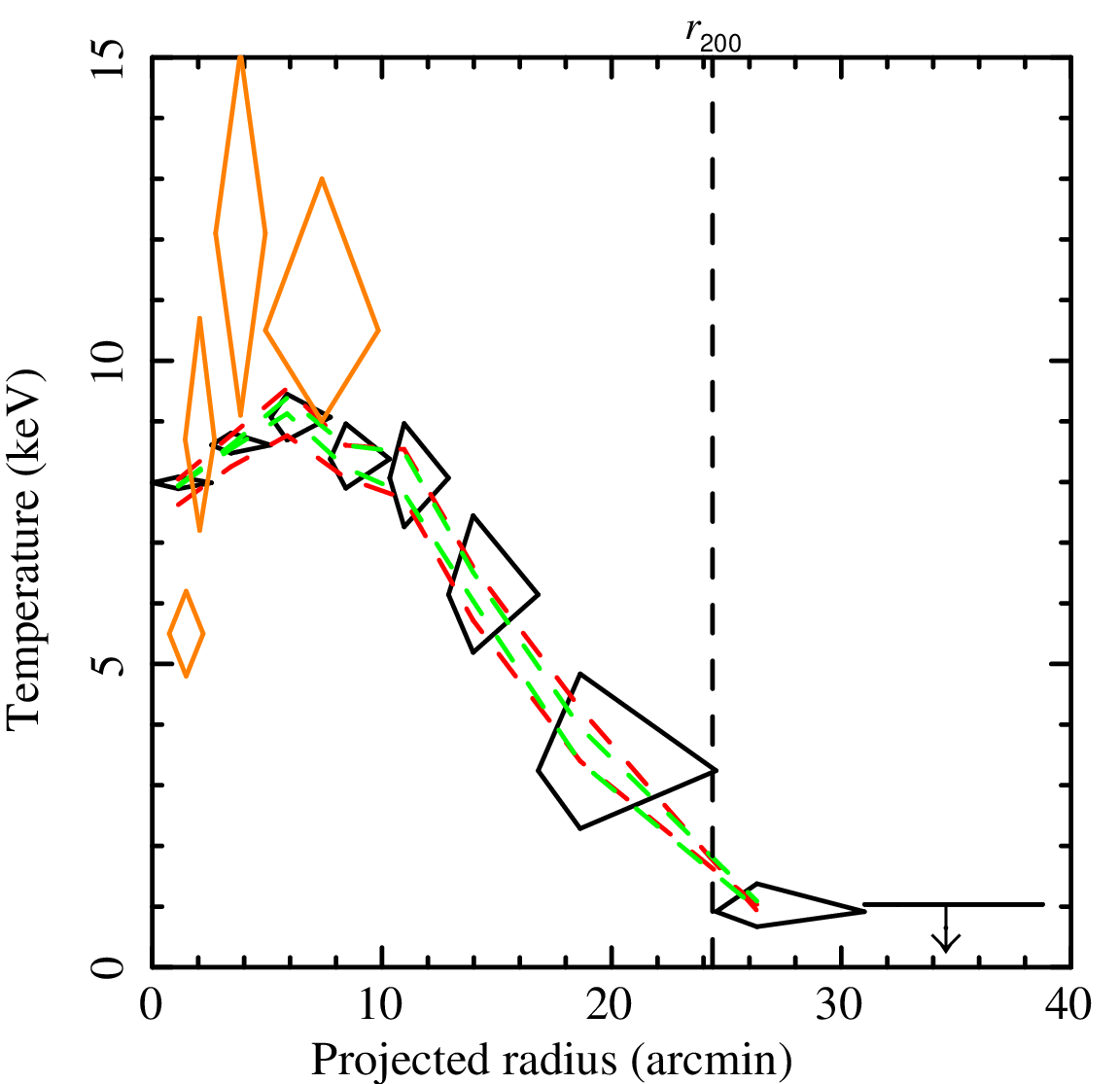}
\end{center}
\end{minipage}\\
\begin{minipage}{1.0\hsize}
(b) Electron density
\\[-.8cm]
\begin{center}
 \includegraphics[angle=-90,scale=0.4]{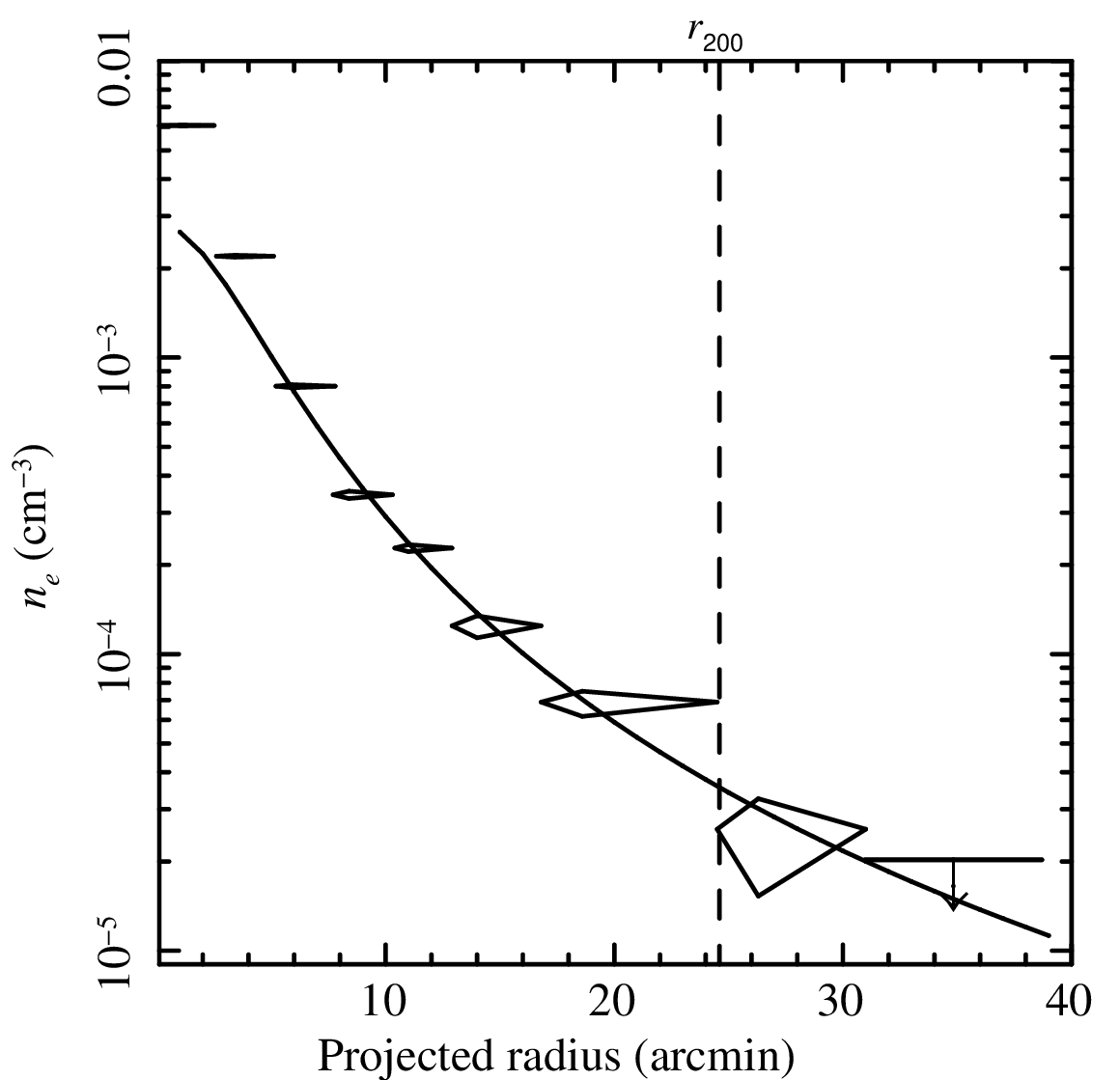}
\end{center}
\end{minipage}
\end{tabular}
\caption{Radial profiles for (a) temperature, (b) 3-dimensional
  electron density.  In the temperature plot, the uncertainty range
  due to the combination of $\pm$ 3\% variation of the NXB level and the
  maximum/minimum fluctuation of CXB is shown by two green dashed
  lines.  In a similar way, the uncertainty range due to the effect of
  contamination on the blocking filter is shown by red lines. Orange
  diamonds show Chandra data from \citet{markevitch00}. }
\label{fig:parameters}
\end{figure}

The resultant radial distribution of $n_{e}$ is shown in
figure~\ref{fig:parameters}~(b), along with the $\beta$-model
profile from ROSAT ($\beta=0.85, r_{c}=\timeform{4.5'}$).  We obtained 
an upper limit of the
electron density in the outermost region (\timeform{31.1'}--\timeform{38.8'}) to be $<
2.0 \times 10^{-5} \rm cm^{-3}$.  One can see that the $\beta$-model
gives a fairly good approximation for the density profile, even though
the temperature shows a large deviation from the isothermal case.

\subsection{Systematic Errors}
\label{sec:err}
We have examined the effect of systematic errors on the derived
spectral parameters.  We considered 3 components for the systematic
errors; namely, the NXB intensity with an error of $\pm 3\%$ \citep{tawa08}, 
fluctuation of the CXB intensity, and contamination on the blocking filter. 
The CXB intensity was estimated in section 3.5 with the fluctuation shown in table~\ref{tab:cxb_fluc}.  
We repeated all the spectral fits by fixing the CXB intensity at the upper and lower boundary values, 
and the resultant parameters are listed in table~\ref{tab:result_off3}.  
The top group shows the parameters with the nominal CXB intensity and contamination thickness.
The parameters in CXBMAX group are for the higher CXB intensity, and
those in CXBMIN are for the lower CXB intensity.  The effects on the
temperature is about 20\%
in the regions near the virial radius.

The other source of the systematic error is the uncertainty in the
amount of contamination on the blocking filter of the XIS instrument.
It showed a gradual increase during the observations of A2142, and the
standard way is to include 10\% error in the thickness of the
contamination layer. The bottom 2 groups in tables
\ref{tab:result_off3} and \ref{tab:result_tcrb} show the
results. Since this error changes the detector response, all the
regions are affected by the same amount. The change of temperature is
10--20\%.

The Galactic foreground emission is direction dependent even though the
spectrum can be approximated by the sum of two thermal models with
temperatures of about 0.1 keV and 0.3 keV\@.  We looked into the results
for two different backgrounds, Offset3 and TCrB regions, which can be
treated as the systematic error due to the spatial variation of the
Galactic emission. As seen in table \ref{tab:result_tcrb}, the effect
becomes larger as the position goes farther from the cluster
center. Near the virial radius, the difference in the ICM temperature
is about 0.5 keV yielding twice different temperatures.

\subsection{Search for WHIM Lines}

\begin{figure}[ht]
\begin{center}
 \includegraphics[angle=-90,scale=0.35]{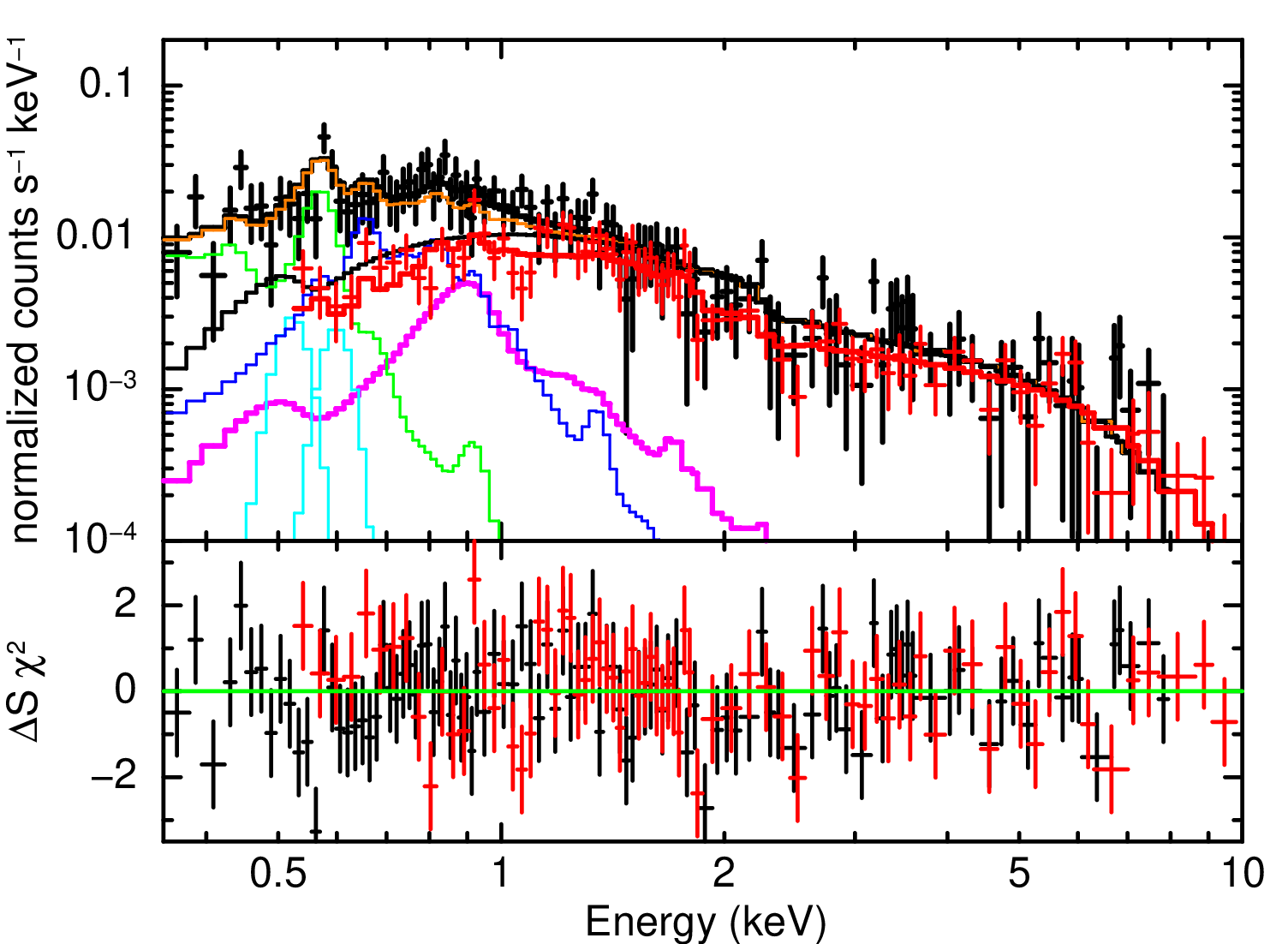}
\end{center}
\caption{
Fits for the constraint on the intensities of the  $I_{\rm
 O\emissiontype{VII}}$
and 
$I_{\rm O\emissiontype{VIII}}$ lines in the \timeform{31.1'}--\timeform{38.8'} region. 
The $I_{\rm O\emissiontype{VII}}$ and $ I_{\rm O\emissiontype{VIII}}$ 
emission lines are shown as a cyan lines, 
and the notations of the other lines are same as shown in figure \ref{fig:fit}.
}
\label{fig:whim}
\end{figure}

We searched for redshifted O lines in the spectra of all the annular
regions, by adding gaussian lines in the spectral fit as shown
in figure~\ref{fig:whim}.  
The energies of the 
O\emissiontype{VII} and O\emissiontype{VIII} 
lines, assumed to have the same redshift of A2142 (0.0909), 
were fixed at 521 eV and 598 eV, respectively.  
We also assumed the O abundance in the ICM to be 0, which gives the highest or most conservative upper limits for the O lines.  
We obtained 2~$\sigma$ upper limits of the O line intensity for all the
regions as shown in table~\ref{tab:whim}. 
The average baryon density in the local universe is $1.77\times 10^{-7}(1+z)^{3}~\mathrm{ cm^{-3}}$ \citep{takei08}, 
or $2.3 \times 10^{-7}\rm\ cm^{-3}$ at $z=0.0909$.  
Based on the observed upper limits of the line intensities, we derived upper limits of the
over-density ($\delta$) of the gas assuming the line-of-sight depth of 2 Mpc
and the temperature of $2\times 10^{6}$ K. 
These assumptions are the same as in \citet{takei08}.  
The results are summarized in table \ref{tab:whim} for the two different Galactic backgrounds separately.
We note that the lowest upper limits, with overdensity less than 280
(assuming background of Offset3) or 380 (assuming background of
TCrB) 
is obtained for the outermost region ($r = \timeform{31.0'}-\timeform{38.7'}$ ).
We also derived upper limits of the overdensity from the measured
electron density.  In sec~\ref{sec:ne}, we showed the 1 $\sigma$ upper-limit of the
electron density to be $< 2.0 \times 10^{-5} \rm\ cm^{-3}$ in the outermost region.
In the range of 2 $\sigma$, the upper limit of the electron density is $< 2.6 \times 10^{-5} \rm\ cm^{-3}$,
corresponding to an overdensity $<~113$.

\begin{table}[ht]
\begin{center}
\scriptsize
\caption{WHIM upper limit}
\begin{tabular}{c|cccc|cccc} \hline
Region &  $ {I_{\rm O\emissiontype{VII}}^{\ast}}$ &${I_{\rm O\emissiontype{VIII}}^{\ast}}$ & $\delta_{\rm O\emissiontype{VII}}$ & $\delta_{\rm O\emissiontype{VIII}}$ & 
$ {I_{\rm O\emissiontype{VII}}^{\ast}}$ &${I_{\rm O\emissiontype{VIII}}^{\ast}}$ & $\delta_{\rm O\emissiontype{VII}}$ & $\delta_{\rm O\emissiontype{VIII}}$ \\
\hline
& \multicolumn{4}{c}{BGD OFF3} & \multicolumn{4}{|c}{BGD TCRB}\\ \hline
Takei et al. & $<$ 1.1    & $<$ 3.0 & $<$ 270 & &&\\ \hline
\timeform{0'-2.5'} & 83.60 & 16.50 & 3311 & 1470 & 85.30 & 17.24 & 3344 & 1503  \\
\timeform{2.5'}-\timeform{5.0'} & 20.28 & 3.79 & 1630 &  705 & 21.85 & 4.51 & 1693 &  769  \\ 
\timeform{5.0'}-\timeform{7.8'} & 3.58 & 1.01 &  685 &  364 & 5.14 & 1.38 &  821 &  425  \\ 
\timeform{7.8'}-\timeform{10.3'} &0.92 & 0.68 &  347 &  299 & 1.48 & 1.06 &  440 &  373  \\ 
\timeform{10.3'}-\timeform{12.9'}&0.79 & 0.30 &  321 &  199 & 1.41 & 0.44 &  429 &  240  \\ 
\timeform{12.9'}-\timeform{16.8'}& 1.34 & 1.82 &  419 &  489 & 2.56 & 2.94 &  579 &  621  \\ 
\timeform{16.8'}-\timeform{24.6'}& 0.73 & 0.64 &  309 &  289 & 1.66 & 1.32 &  466 &  416  \\ 
\timeform{24.6'}-\timeform{31.1'} & 0.60 & 0.62 &  281 &  284 & 1.66 & 1.46 &  466 &  437  \\ 
\timeform{31.1'}-\timeform{38.8'}& 0.57 & 1.01 & 275 & 363 & 1.08 & 1.32 & 377 & 415  \\ \hline
\multicolumn{9}{l}{
$\ast:10^{-7} \rm ~ph~cm^{-2}~s^{-1}~arcmin^{-2}$ with 2$\sigma$ upper limit.
}
\end{tabular}
\label{tab:whim}
\end{center}
\end{table}

\section{Discussion}\label{sec:discussion}

Suzaku performed four pointing observations in A2142 and its outside
regions along the merger axis.  The temperature was measured out to
the virial radius ($r_{200}\sim 2.5$ Mpc) for the first time.  The ICM
temperature was found to drop from about 9 keV around the center to
about 3 keV at $r_{200}$.  We detected no significant signal from the WHIM
and set upper limits of its over-density.  We will attempt to evaluate
the cluster properties (temperature, electron density, and entropy)
and discuss their implications.

\subsection{Temperature Profile}
Some numerical simulations \citep{ettori04, borgani04} predicted that
the intracluster gas temperature drops to about 50\% of the central
temperature around $r_{200}$ under hydrostatic equilibrium.  
Those results reproduced observed temperature profiles to about $0.5r_{200}$ \citep{degrandi02,vikhlinin05,pratt07}.  
In particular, recent XMM-Newton~\citep{pratt07} results showed temperature profiles out to $0.8r_{200}$.

\begin{figure}[ht]
\begin{center}
 \includegraphics[angle=-90,scale=0.35]{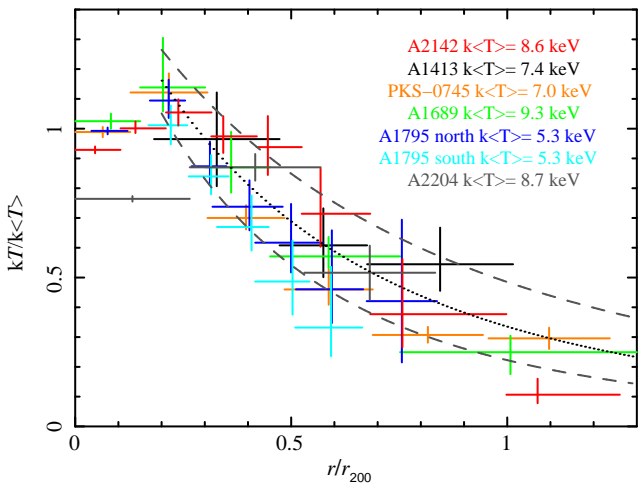}
\label{fig:kt}
\end{center}
\caption{ Scaled projected temperature profiles. 
The profiles have been normalized to the mean temperature.
The $r_{200}$ derived from \citet{henry09}.  Dotted line show simulation result~\citet{burns10}. 
Two gray dashed lines show standard deviation.
}
\label{fig:kt-all}
\end{figure}

We compare here the observed temperature profile of A2142 with the pro?le of other clusters that
 have been observed with Suzaku.  
Table \ref{tab:kt-sum} shows the list of clusters for which Suzaku measured the temperature out to $r_{200}$. 
Their temperature profiles are shown in figure~\ref{fig:kt-all}.  
All the temperatures are normalized by the flux-weighted average of each clusters.  
The results clearly show a systematic drop of the temperature by a factor of 3-5 from the center to $r_{200}$.

Among these clusters, A1689 shows a significant directional difference
in the sense that the temperature drop is small along the filament direction \citep{kawaharada10}.  
Another system A1413 also indicates a somewhat flatter temperature profile, and the measured direction 
is along the longer axis of its X-ray elongation \citep{hoshino10}.

\citet{burns10} discuss a non-equilibrium effect on the temperature profile
based on N-body + hydrodynamic simulations.  Those simulations
indicate cluster temperature decline by factor of $\sim 3$ at
$r_{200}$, consistent with the feature observed for two nearby
clusters, PKS 0745-191 and A 1795 \citep{george08,bautz09}.
\citet{burns10} approximated the average temperature profile by a function,
\begin{equation}
\frac{T}{T_{\rm avg}}=A\left[1+B\left({\frac{r}{r_{200}}}\right)\right]^{\beta}.
\label{eq:burns}
\end{equation}
They obtained the best-fit values as $A=1.74\pm 0.03,\ B=0.64\pm 0.10, \
\beta=-3.2\pm 0.4$ for the two cluster data.  In figure
~\ref{fig:kt-all}, we plotted model prediction temperature profile.
Black dotted curve shows the best-fit temperature profile, and dashed
lines show $1\sigma$ error range reported by \citet{burns10}.  The
relation of \citet{burns10} represents the temperature profiles for the
6 clusters fairly well.  This approximate ``universal'' temperature
profile suggests that clusters generally hold self-similar relation
even near $r_{200}$, where some temporary effects caused by infalling
matter may be seen.  Note that the temperature drop in the filament
direction of A1689 (not shown in Fig~\ref{fig:kt-all}) is flatter than
the average A1689 profile, suggesting a very efficient heating going
on in the filament direction as compared with typical clusters.

\begin{table}[h]
\scriptsize
\begin{center}
\caption{Cluster samples and those properties}
\begin{tabular}{ccccccc} \hline \hline
Cluster	& $z$ & Ref.\  	& $k\langle T\rangle$  & $r_{200}$ \\ 
			&		&					&[keV]	&[Mpc](arcmin) \\ \hline
Abell 2142 	& 0.090 	& This work 			& 8.6	& 2.46 (24.6) \\
PKS 0745-191  & 0.103 	&\cite{george08} 		& 7.0 	& 2.21 (19.6) \\
Abell 2052 	& 0.036 	&\cite{tamura08} 		& 3.2 	& 1.54 (36.7) \\
Abell 2204 	& 0.152 	&\cite{reiprich09} 		& 8.7	& 2.40 (13.2) \\
Abell 1795 	& 0.063   &\cite{bautz09} 		&5.3 	& 1.96 (26.9) \\
Abell 1413 	& 0.143 	&\cite{hoshino10}		& 7.4 	& 2.24 (14.8) \\
Abell 1689 	& 0.183   & \cite{kawaharada10}  & 9.3  	& 2.44 (13.3) \\
Perseus  	&0.018	&\cite{simionescu11}	 &   6.5 		& 2.22 (103.1)\\ \hline
\end{tabular}
\label{tab:kt-sum}
\end{center}
\end{table}

All the clusters in figure~\ref{fig:kt-all} are morphologically
relaxed systems, even though all of them show an elliptical shape to
some extent.  Therefore, we may regard the model temperature profile
to be a typical feature for clusters having relaxed morphologies.  We
note that the present A2142 profile shows a smooth temperature decline
with radius, even though it is a merger cluster characterized by clear
cold fronts within $\timeform{3'}$ ($0.11 r_{200}$) from the center.

We note that when we fit the temperature profiles with the above
formula \ref{eq:burns}, A2142 data are characterized by $\beta
=-5.8\pm 0.8$, which indicates a steepr slpe than the average, $\beta
= -3.2 \pm 0.4$.  On the other hand in A1689, the filament direction
shows a flatter temperature profile \citep{kawaharada10} than the
average curve. It has been discussed that A1689 still holds the
heating feature caused by recent matter infall.  The same process may
be working in A1413 which also shows somewhat flatter temperature
distribution \citep{hoshino10} even though the presence of filament is
not clear in the observed direction.

Even though flatter temperature profiles are regarded as a sign of
recent matter infall, steeper temperature curves such as seen in
A~2142 and abrupt temperature drops, as seen in A~3667 and Virgo Cluster
can also be a sign of recent heating \citep{finoguenov10, urban11}. 
This point will be discussed further in the next section. 
Therefore, simple steepness of the temperature profile seems to carry 
somewhat degenerate information about the cluster evolution.

\subsection{Entropy Profiles}

The entropy of ICM is used as an indicator of the energy acquired by
the gas. We will refer to ``entropy'' of the ICM by $ K = kT
n_e^{-2/3}$ following the recent convention.  Numerical simulations
indicate that a self-similar growth of clusters commonly show entropy
profiles approximated by $r^{1.1}$ up to $r_{200}$, excluding the cool
core region~\citep{voit03}.  Recent XMM-Newton results on the entropy
profiles of 31 clusters showed a median slope of 0.98 out to about
$r_{500}$ which is approximately $0.5 r_{200}$ \citep{pratt10}.  The
slope also showed positive correlation with the average temperature.
They also found that morphologically disturbed clusters show a large
scatter (0.5--2.0) in the slope.  Suzaku has extended the entropy
measurement close to $r_{200}$ for several clusters, and showed a
flattening or even a decrease at $r\gtrsim 0.5r_{200}$ as mentioned
earlier.  Discussion has been made that the ICM may not be under
equilibrium in the cluster outer regions \citep{hoshino10}.

\begin{figure}[ht]
\begin{tabular}{c}
\begin{minipage}{1\hsize}
(a) Entropy
\\[-.7cm]
\begin{center}
 \includegraphics[angle=-90,scale=0.3]{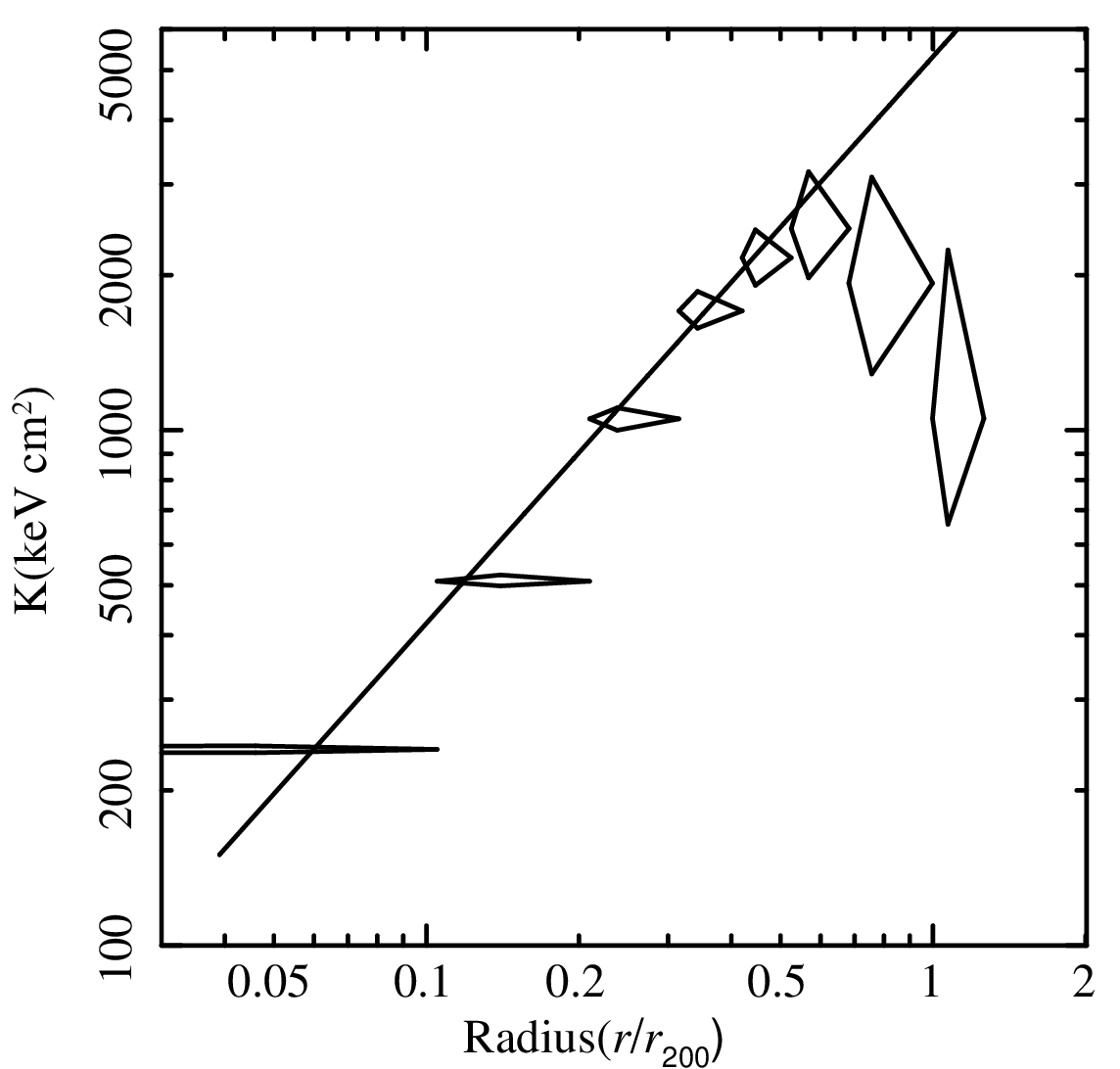}
\end{center}
\end{minipage}\\
\begin{minipage}{1\hsize}
(b) Ratio to $r^{1.1}$ consistent with $T_{e}/T_{gas}$
\\[-.7cm]
\begin{center}
 \includegraphics[angle=-90,scale=0.3]{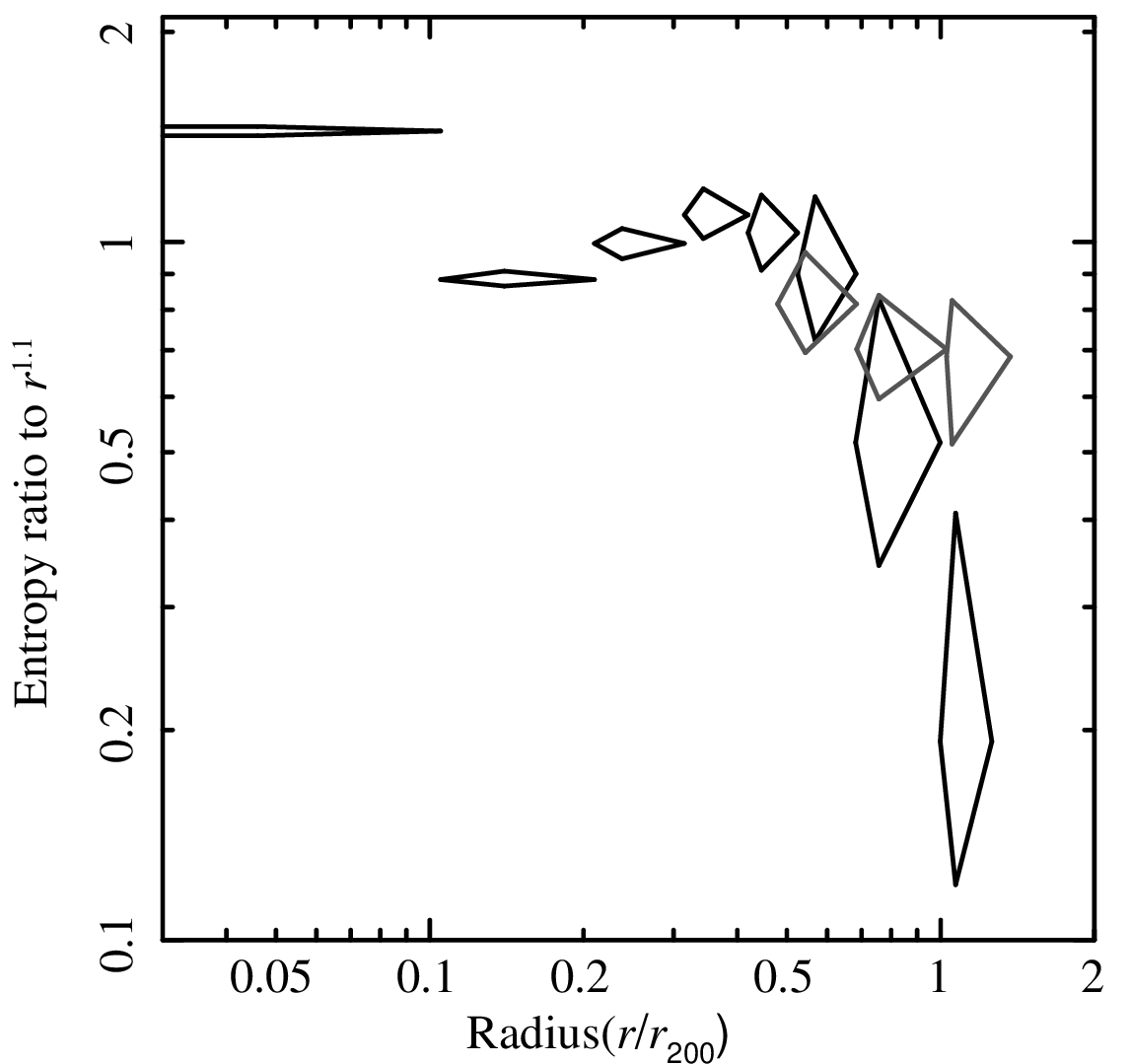}
\end{center}
\end{minipage}\\
\begin{minipage}{1\hsize}
(c) Equilibrium time
\\[-.7cm]
\begin{center}
 \includegraphics[angle=-90,scale=0.3]{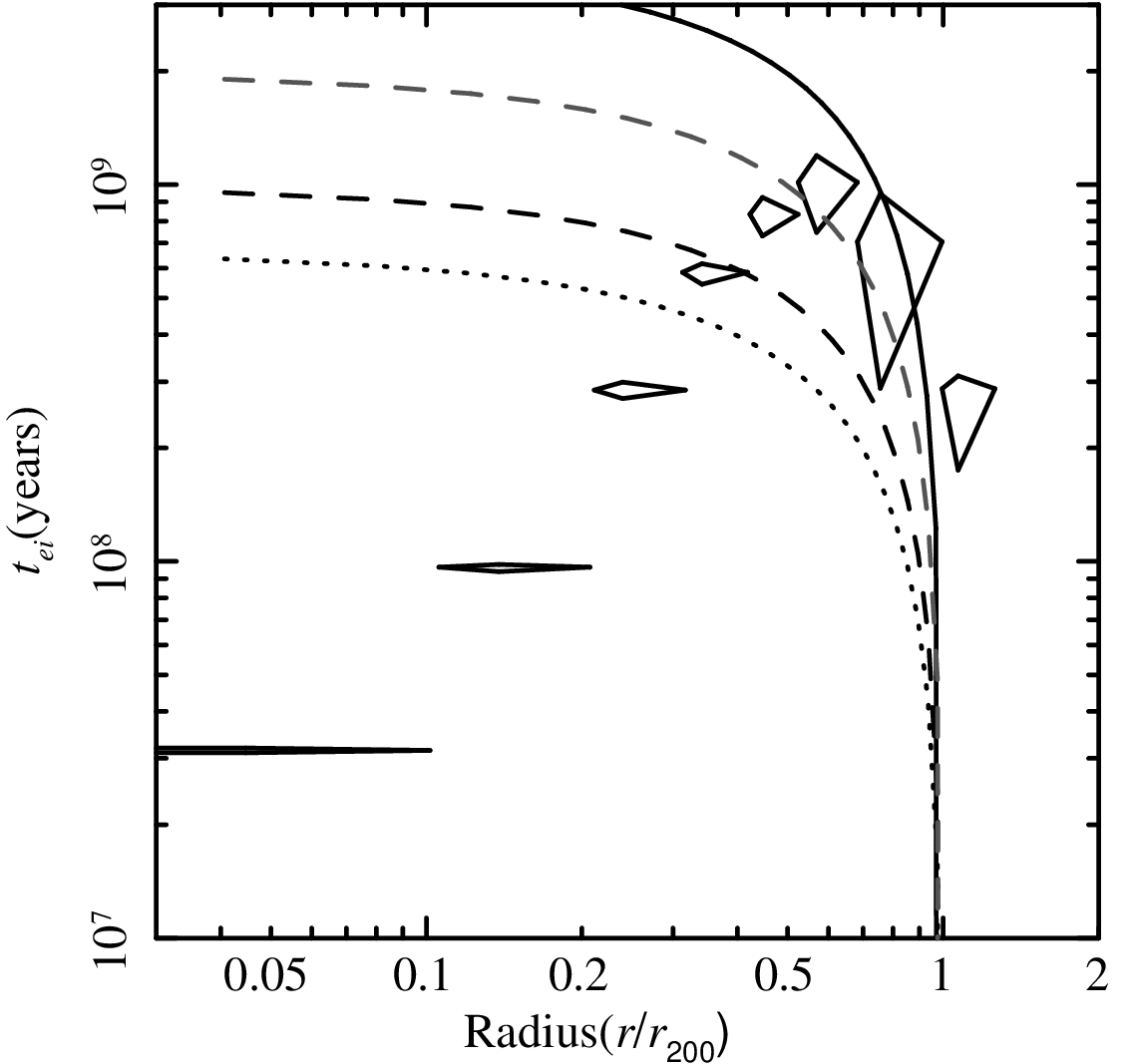}
\end{center}
\end{minipage}
\end{tabular}
\caption{ (a) Entropy profile: Diamonds show Suzaku results, and solid
  straight line shows the universal trend of the entropy curve $\sim
  r^{1.1}$, reported by \citet{voit05}.  (b) Entropy ratio relative to
  the $r^{1.1}$ profile.  Gray diamonds show the result of A1413
  \citep{hoshino10} (c) $t_{ei}$ (diamonds) as a function of radius,
  compared with estimated $t_{\rm elapsed}$ (solid line).  Different
  lines show different shock speeds (gray dash:400 km s$^{-1}$, dash:
  800 km s$^{-1}$, dot: 1200 km s$^{-1}$).  }
\label{fig:entropy}
\end{figure}

Figure~\ref{fig:entropy} shows the entropy profile of A2142 based on
the present Suzaku measurements.  The solid line indicates the slope
of 1.1.  The entropy slope is consistent with this value in the radius
range $0.1-0.4 r_{200}\ (\timeform{2.5'}-\timeform{10'})$.  The slope
becomes flatter at $r > 0.4 r_{200}$ and goes negative near $r_{200}$.
This feature is more clearly seen in figure \ref{fig:entropy}(b),
which shows the ratio to the $r^{1.1}$ curve which can be fitted for
the inner profile, $r < 0.4 r_{200}$. Including the previous results
for PKS0745-191 and A1413, flattening of the entropy profile in such
outer regions looks to be a common feature.

As for the cause of the entropy drop relative to the predicted $r^{1.1}$
relation, two possibilities have been pointed out so far. One is the
non-equilibrium effect \citep{hoshino10,burns10} and the other is the
clumpiness \citep{simionescu11}, both working only in the outer
regions.  These two conditions are not mutually exclusive but may be
closely related with each other. 

Regarding the effect of clumpiness, \citet{nagai11} examined with
numerical simulations how much clumping in accreting gas could cause
overestimation of gas density since the emissivity scales as density
squared.  This density overestimation gives lower entropy values,
hence resulting in a flattening around $r_{200}$.  Results of
\citet{nagai11} show that the entropy drops by 10-15\% from the true
value at around $r_{200}$. In the A2142 case, the observed entropy
curve starts to flatten around $r_{500}$ and decreases to less than
40\% of the $r^{1.1}$ extention at $r_{200}\ (\sim \timeform{25'})$.
This large suppression of the entropy seems too large according to the
simulation. Therefore, clumping is not likely to be the only or the
dominant source of the entropy suppression.

\subsection{Ion-Electron Relaxation}
\label{sec:nei}
The entropy profile of A2142 shows a flatter slope than the predicted
one based on numerical simulations, and even a negative slope is seen
in the outermost region.  \citet{kawaharada10} showed a flat entropy
profile in the filament direction of A1689 and discussed effects from
surrounding environments.  If we make a simple assumption that the
entropy profile of ICM should follow $r^{1.1}$~\citep{voit03} under an
equilibrium, the present result for A2142 suggests that electrons in
the cluster outer region is not in the equilibrium.

\citet{hoshino10} discussed the possible difference between ion and
electron temperatures in the outer low-density region, where the
relaxation time scale is inversely proportional to particle
density. We follow their estimation and look into the behavior of ion
($T_i$) and electron ($T_e$) temperatures. The following assumptions
are made.
\begin{enumerate}
\item Ions are initially heated through accretion shocks at $r_{200}$.
\item Ions achieve thermal equilibrium with a timescale of ion-ion
  relaxation, $t_{ii}$, after this heating.
\item Thermal energy is transferred from ions to electrons through
  Coulomb collisions. This process takes a timescale of $t_{ei}$ which
  is much longer than both $t_{ii}$ and the electron-electron
  relaxation time $t_{ee}$.
\end{enumerate}

The relation between $T_i$ and $T_e$ can be estimated in the following
way after \citet{hoshino10}.  We look into a position-dependent time,
$t_{\rm elapsed}$, which is a rough measure of the time elapsed from
the shock heating.  We assume that the shock heating takes place at
the virial radius and that the shock wave propagates with a constant
speed (400, 800, 1200 km s$^{-1}$) in ICM toward inner regions.  This
$t_{\rm elapsed}$ can be compared with the equilibration timescale
$t_{ei}$, and we may assume that electrons are fully heated up when
$t_{\rm elapsed} > t_{ei}$.  In figure~\ref{fig:entropy}(c), we
compare $t_{\rm elapsed}$ with $t_{ei}$ as a function of radius.
Each curves show the position-dependent $t_{\rm elapsed}$, assuming
that the gas falls through either free-fall or constant velocities.
The equilibration timescale $t_{ei}$ is significantly longer than
$t_{\rm elapsed}$ in the cluster outskirts at $r > \timeform{20'}$,
suggesting that the ion temperature can be higher than the electron
temperature. This feature is essentially the same in A~1413 as
reported by \citet{hoshino10}.

Theoretical studies of non-equilibrium ionization state and an electron-ion two-temperature structure of ICM in
  merging galaxy clusters have been carried out  \citep{rudd09,akahori10}. 
  They show that in merging clusters,$T_e$
  is lower than the average temperature by 20--30\%.  Following
  \citet{hoshino10}, we estimate the possible deviation of $T_e$ by
  assuming that the average temperature is given by the entropy and
  density as $kT_{\rm gas}=Kn^{2/3}$, and further that the entropy $K$
  follows the power-law profile $\propto r^{1.1}$ to $r_{200}$
  \citep{voit03}.  Figure~\ref{fig:entropy}(b) shows thus estimated
  ratio $T_e/T_{\rm gas}$.  Previous A1413 results by
  \citet{hoshino10} are shown for compariton.  A2142 values are lower
  than A1413 at $r_{200}$, which may reflect stronger suppresion of
  $T_e$ in merging clusters.  The $T_{e}/T_{\rm gas}$ value in A2142 is
  $0.51^{+0.31}_{-0.17}$ at $r_{200}$, which is in good agreement with
  the theoretical result for unrelaxed clusters \citep{rudd09}. 

Even though this feature suggests that $T_e$ is
  substantially lower than the average gas temperature around
  $r_{200}$, we have to note that the temperature profile of A2142
  agrees with those for other relaxed clusters as seen in
  figure~\ref{fig:kt-all}. This may suggest that the entropy drop is
  rather due to high $n_e$ around $r_{200}$. However, as shown in
  figure \ref{fig:parameters} (b), the observed $n_e$ profile also
  follows a smooth $\beta$-model and no peculiar hump is seen in the
  outer region. It may be that the outer region of A2142 has not
  experienced strong merger recently, but then the entropy
  ``saturation'' in the outer region will have to have a more
  universal origin. We certainly need to look into this problem with a
  wider range of cluster sample regarding the cluster size and
  morphology. 

\subsection{Electron Density Profile}
The electron number density profile in figure \ref{fig:parameters}
shows a decrease down to $\sim 10^{-5}$ cm$^{-3}$ around the virial
radius.  ROSAT study showed that the electron density profile could be
fitted well except for the cluster center with the $\beta$-model,
after modification for the cluster ellipticity \citep{henry96}.  In
the outskirts of relaxed clusters, many systems indicate a density
profile of $n_{e}\propto r^{-2.2}$ ~\citep{zhang06}.  Also, $n_{e}
\propto r^{-1.8\pm0.28}$ is obtained for the REXCESS sample
\citep{croston08}

The measured A2142 density profile is fitted with a power-law model
with an index of $-1.98 \pm 0.13$ for the entire radial range and
$-2.53 \pm 0.25$ in the outskirts only ($r > 0.5 r_{200}$).  These
results agree with the previous studies for other clusters, and with
the predicted profile of $n_{e} \propto r^{-2.5}$ based on the
$\beta$-model ($\beta=0.85$) in the outer region.

Recently, \citet{eckert11} pointed out a factor of
  about 3 discrepancy in the surface brightness in the outer region of
  PKS 0745-191 between Suzaku and ROSAT results. They argue that it is
  most likely to be caused by an incorrect subtraction of the Galactic
  emission for the Suzaku data.  As for A2142, ROSAT PSPC observed its
  emission to $r_{200}$ \citep{henry96} and showed $n_e \sim 5\times
  10^{-5}$ cm$^{-3}$ at $r_{200}$.  The present $n_e$ profile, shown
  in figure \ref{fig:parameters}, indicates a lower but consistent
  value with the ROSAT one within a factor of $\sim 1.5$. Therefore,
  our subtraction of the Galactic background in the nearby sky region
  is considered to give reliable results.

\subsection{Mass Estimation to $r_{200}$}
We estimate the gravitational mass of A2142 to $r_{200}$ based on the observed temperature and density profiles. 
Here, we assume hydrostatic equilibrium and spherical symmetry, and calculate the gravitational
mass within 3-demensional radius $R$ with the following formula \citep{fabricant80},

\begin{equation}
M_{R}=-\frac{R^2}{\rho_gG}\frac{dP_g}{dR}=-\frac{kTR}{\mu m_p G}
\left(\frac{d \ln \rho_g}{d \ln R}+\frac
{d \ln T}{d \ln R}\right),
\end{equation}
where $G$ is the gravitational constant, and 
$\mu (\approx 0.6)$ is the mean molecular weight of gas.

\begin{figure*}[ht]
\begin{tabular}{cc}
\begin{minipage}{0.5\hsize}
(a) Total integrated gravitational mass
\\[-0.5cm]
\begin{center}
 \includegraphics[angle=-90,scale=0.4]{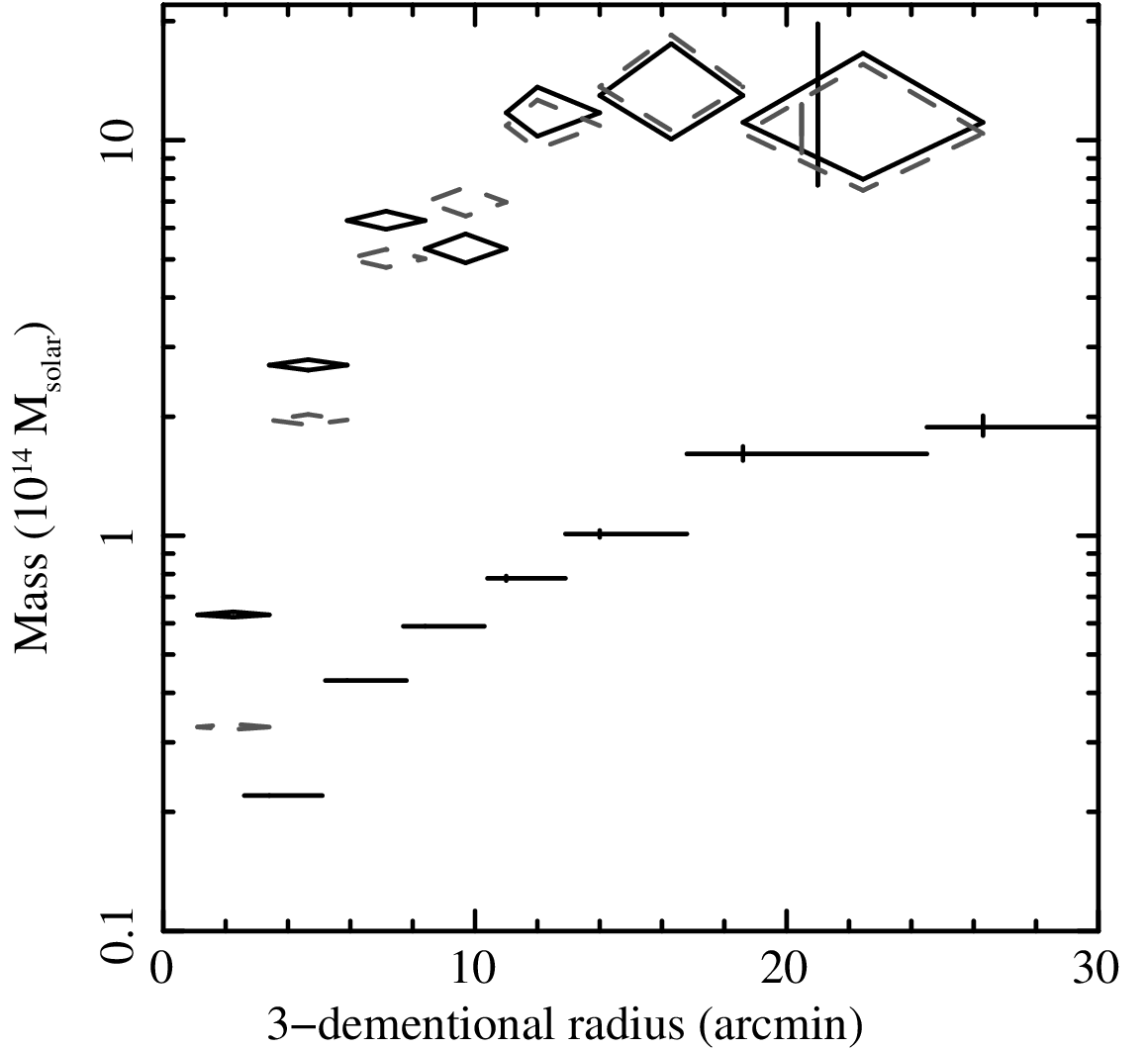}
\end{center}
\end{minipage}
\begin{minipage}{0.5\hsize}
(b) Differential mass density
\\[-0.5cm]
\begin{center}
 \includegraphics[angle=-90,scale=0.4]{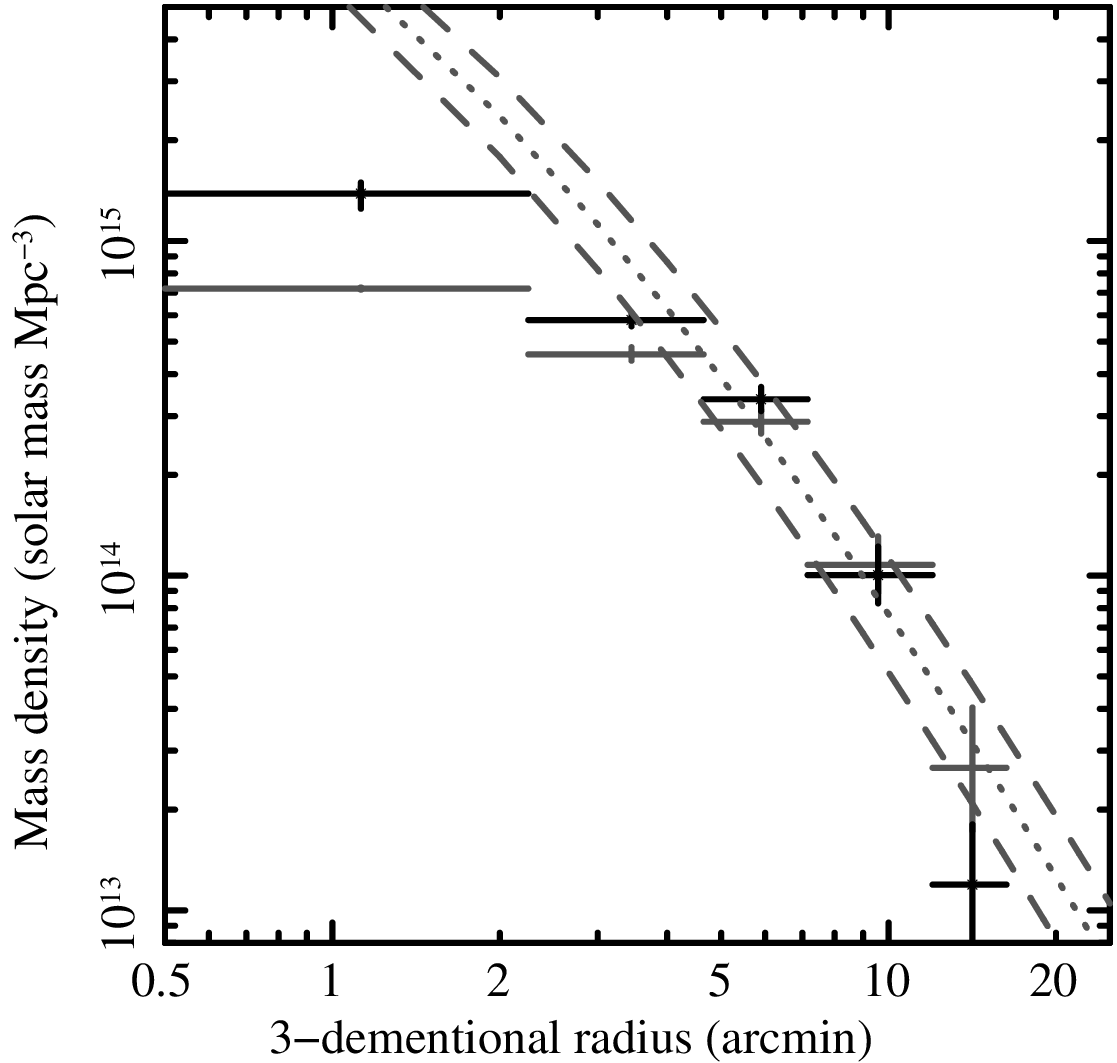}
\end{center}
\end{minipage}
\end{tabular}
\caption{
(a) Total integrated gravitational mass $M_{<R}$ profile of A2142. 
Black diamonds show gravitational mass profile estimated from the Suzaku data.  
Gray dashed diamonds show a profile assuming the electron density to follow $\beta$ model distribution with $\beta = 0.85$. 
Black and gray vertical lines show masses estimated by weak-lensing analysis summarized in table~\ref{tab:mass}~\citep{okabe08,umetsu09}.
Black cross show gas mass profile estimated from the observed electron density profiles.
(b) Same as (a), but for differential mass density. 
We ignore the data around \timeform{10'},
because it gives an unphysical negative value, 
Dotted curve show NFW-model results by~\citet{umetsu09}.
Two dashed lines show standard deviation.
}
\label{fig:mass}
\end{figure*}

In figure \ref{fig:mass}, solid diamonds show the gravitational
mass of A2142 based on the observed temperature and density profiles.
Two solid vertical lines (black and gray) indicate the mass derived
from the previous weak lensing studies which are summarized in
table~\ref{tab:mass}~\citep{okabe08,umetsu09}.
Dashed-line diamonds show gravitational mass using the $\beta$-model
density profile together with the observed temperature gradient.  The
resultant mass agrees well with the previous studies around the virial radius.

We compare the parameters of an NFW profiles, defined as
\begin{equation}
\rho=\frac{\rho_{s}}{(r/r_{s})(1+r/r_{s})^{2}}
\end{equation}
where $\rho_{s}$ is central density parameter and $r_{s}$ is the
scaled radius.  Figure~\ref{fig:mass}(b) show the resultant values of
differential mass density ($c=2.8_{-1.2}^{+1.0}$), which reflect the dark matter potential.  Our
differential mass density profile is consistent with the previous
weak-lensing result which shows $c=4.26^{+0.71}_{-0.63}$
\citep{umetsu09}.  

We can evaluate the contribution of non-thermal pressure by comparing
the X-ray and the weak-lensing masses.  The X-ray to lensing mass
ratio is $0.90^{+0.48}_{-0.35}$, which indicates that, by taking the
lowest boundary value, the maximum non-thermal pressure can be 45\%.
As shown in \S\ref{sec:nei}, we discussed the possible difference
between ion and electron temperatures as the cause of the entropy drop
from the $r^{1.1}$ relation.  
In order for the entropy profile to follow the $r^{1.1}$ relation after electron
temperature reaches the ion temperature, ion temperature should have
higher temperature than electrons.
The ratio of electron to ion temperature
is estimated to be $0.51^{+0.31}_{-0.17}$ around the virial
radius (see fig\ref{fig:entropy}b: this value also indicates
the  ratio of electron to ion temperatures).
Then, since the gravitational force is balanced with the
sum of electron and ion pressure in hydrostatic equilibrium,
the gravitational mass would rise roughly by the same factor.
Although this leads to poorer match between X-ray and lensing mass,
they are still barely consistent within statistical and systematic
uncertainties.

We note that non-thermal pressure can make additional contribution to
the cluster mass estimation. Dynamical effects such as turbulence and
bulk motion of ICM are not yet measured, but can give significant
effects in the cluster outer regions.  Some numerical simulations
\citep{nagai07,piffaretti08} predicted that such non-thermal pressure
could add up 
15-30\% of the cluster mass.

We also looked into the gas mass distribution using
  the obtained electron density profile and showed the results in
  figure \ref{fig:mass}(a) with black crosses.  Here, the gas mass
  does not include the stellar mass.  The resultant gas mass fraction
  at $r_{200}$ is $14.4^{+7.1}_{-4.1}\%$. This agrees with the
  expected hot-gas fraction (15\%) in the universe \citep{komatsu11},
  and consistent with the previous result, $\sim18\%$ at $22'.8$,
  incorporating the weak lensing and Sunyaev and Zel'dovich
  observations \citep{umetsu09}.  This indicates that, in A2142, there
  is no strong need to invoke the gas clumpiness to account for the
  gas mass fraction as in the Perseus cluster \citep{simionescu11}.

\begin{table}[h]
\caption{Mass estimation of A2142}
\centering
\begin{tabular}{ccccccccc}\hline
Reference 		&$r$ &$M_{200}$	\\
				& 	arcmin	&$10^{14} M_{\odot}$ \\ \hline
\cite{okabe08}		  &22.2	&$13.7\pm6.0$\\
\cite{umetsu09}		 &22.2	&$12.3_{-2.0}^{+3.0}$\\
This work (Observed gas density profile)		 &22.5$\pm$3.9	&$11.1_{-3.1}^{+5.5}$ \\ 
This work ($\beta$-model gas density profile)		 &22.5$\pm$3.9	&$10.4_{-2.9}^{+5.2}$ \\ 
\hline
\end{tabular}
\label{tab:mass}
\end{table}

\section{Summary}
We observed Abell~2142 in the direction of the possible merger axis
with Suzaku and detected the ICM emission up to the virial radius $r_{200}$ (2.5 Mpc $\sim \timeform{24.6'}$).  
We derived radial profiles of temperature, electron density, and entropy, 
and compared these properties with the previous results for other relaxed clusters.  
We summarize the main features of A2142 as follows;

\begin{itemize}
\item The ICM temperature gradually decreases toward the outer region
   from about 10 keV at $0.2 r_{200}$ to about 4 keV at $r_{200}$.
\item The temperature profile in the outer region of A2142 agrees 
well with the results of other clusters observed by Suzaku.
\item The average temperature profile for different clusters can be
  described by the formula by \citet{burns10} up to $r_{200}$,
  suggesting that non-thermal pressure support is significant in the outer regions.
\item The electron density profile decreases down to $\sim 10^{-5}\
  \rm cm^{-3}$ at the virial radius and well agrees with the $\beta$ model with $\beta = 0.85$.
\item The entropy profile within about $0.4 r_{200}$ follows
  $r^{1.1}$, predicted by the accretion shock heating model.  The
  profile becomes flatter and finally shows a negative slope around
  $r_{200}$, suggesting significant deviation from the equilibrium
  condition.
\item Based on the temperature and entropy profiles, and required
      relaxation time, we discuss that
  $T_i$ is likely higher than $T_e$ in the outer region of A2142.
\item The derived mass profile is in agreement with the weak-lensing
  mass.  The difference between X-ray and lensing masses allowed by
  the error can explain the contribution of the non-equilibrium
  effect.
\end{itemize}

Resent Suzaku results on the temperature and entropy measurements to
$r_{200}$, including the A2142 ones, jointly suggest that ICM in
cluster outskirts is deviating from thermal equilibrium in the sense
that the electron temperature could be significantly lower than the
ion temperature.  This suggests that the ICM around $r_{200}$ has
experienced bulk motions and/or turbulence within a time scale of
about $10^9$ yr.  Such gas motions with a velocity of a few hundred to
thousand km s$^{-1}$ can be observable in future by X-ray
microcalorimeters, such as by SXS instrument on ASTRO-H
\citep{mituda10} which has 20--30 times higher energy resolution than
CCD instruments.  
Since the FOV of SXS is rather
  small ($\timeform{3'}\times\timeform{3'}$), expected counts from Fe-K
  lines are typically around 10 and 60 at $r_{200}$ and $0.5 r_{200}$,
  respectively, for a 300 ksec exposure. Therefore, long exposures for
  selected good targets with ASTRO-H will be needed to show the
  dynamical features appearing in the cluster outer regions for the
  first time.

\bigskip We are grateful to N. Okabe and K. Umetsu for information
about thier Suzaku data and N. Okabe providing SDSS galaxy
distribution data.  H. A. was supported by Grant-in-Aid for JSPS
Fellows (22$\cdot$1582) and the MEXT program `` Support Program for
Improving Graduate School Education''.  NO acknowledges financial
support from Grant-in-Aid for Scientific Research No.22740124.
 
\appendix
\section{Point Source Analysis}
\label{sec:appa}
\begin{table*}[ht]
\begin{center}
\caption{Informations of point source in two XMM  and Suzaku observations of A2142 center, OFFSET1, OFFSET2 OFFSET3. }
\begin{tabular}{ccccccccccccccccccccccccc} \hline
	& 	&	\multicolumn{3}{c}{XMM-Newton(MOS1+MOS2)} 	&\multicolumn{3}{c}{Suzaku}\\
No.$^{\ast}$	& ($\alpha, \delta$)	$^{\dagger}$&Photon Index &Flux$^{\ddagger}$ &$\chi^{2}$/d.o.f	&Photon Index &Flux$^{\ddagger}$ &$\chi^{2}$/d.o.f\\
\hline
1 & (239.401, 27.485) & $ 1.44 ^{+0.22} _{-0.21} $ & $ 1.28 ^{+0.72} _{-0.34}  $ &  108.9/ 63 & $ 1.96 ^{+0.29} _{-0.23} $ & $ 1.02 ^{+0.37} _{-0.30}  $  & 53.0 / 52 \\ 
2 & (239.393, 27.287) & $ 1.42 ^{+0.63} _{-0.68} $ & $ 0.31 ^{+0.56} _{-0.31}  $ &  35.6 / 23 & $ 1.69 ^{+0.14} _{-0.14} $ & $ 1.40 ^{+0.52} _{-0.48}  $  & 59.0 / 53 \\ 
3 & (239.334, 27.275) & $ 2.21 ^{+0.64} _{-0.50} $ & $ 1.27 ^{+0.46} _{-0.18}  $ &  34.0 / 31 & $ 2.08 ^{+0.15} _{-0.13} $ & $ 1.88 ^{+0.46} _{-0.42}  $  & 75.0 / 53 \\ 
4 & (239.283, 27.366) & --- & ---&  --- / --- & $ 1.64 ^{+0.23} _{-0.18} $ & $ 2.32 ^{+0.59} _{-0.59}  $  & 74.0 / 52 \\ 
5 & (239.532, 27.351) & --- & --- &  --- / --- & $ 0.94 ^{+0.54} _{-0.74} $ & $ 2.28 ^{+6.51} _{-1.79}  $  & 58.0 / 51 \\ 
6 & (239.295, 27.605) &---& ---&  --- / --- & $ 1.89 ^{+0.19} _{-0.18} $ & $ 0.89 ^{+0.19} _{-0.18}  $  & 70.0 / 54 \\ 
8 & (238.922, 27.686) &---& ---&  --- / --- & $ 1.93 ^{+0.10} _{-0.10} $ & $ 3.23 ^{+0.41} _{-0.39}  $  & 60.0 / 54 \\ 
9 & (239.012, 27.773) &---& ---&  --- / --- & $ 1.63 ^{+0.15} _{-0.14} $ & $ 1.65 ^{+0.28} _{-0.26}  $  & 52.0 / 54 \\ 
\hline 
\multicolumn{3}{l}{
$\ast$:Serial number for point source.
} &
\multicolumn{3}{l}{
$\dagger$:Position of  the point source.
} \\
\multicolumn{3}{l}{
$\ddagger$:The 2.0--10.0 keV flux in units of  $10^{-13}$ \fluxunit
} 
\end{tabular}
\end{center}
\end{table*}

As for the point-source subtraction, we first analyzed the XMM-Newton data (Observation ID=0111870101, 0111870401) 
in which faint sources were resolved better than the Suzaku data.   
The data covered up to the virial radius ($\sim 2.5$ Mpc).  
We used ${\it wavdetect}$ tool in CIAO (CIAO version:4.0.1) to detect point sources
and used surround annular region for background subtraction.
We summed MOS1 and MOS2 spectra to increase statistics, 
and fitted by $pegpower law$ model which offered photon index and flux in selected energy band.
In the outer region ($r > \timeform{25.6'}$), we removed the sources from the Suzaku data which were selected by eye.  
In the XMM-Newton case, the source extraction radius is $\timeform{30''}$,
and the surrounding background ring in estimating the flux is defined by $\timeform{30''}-\timeform{60''}$, respectively.  
In the Suzaku case, those are $\timeform{1'}$ and $\timeform{1'}-\timeform{2'}$ , respectively.
We found 3 sources in the XMM-Newton data whose fluxes were 
higher than $3\times10^{-14}$ erg cm$^{-2}$ s$^{-1}$ in the energy range 2 -- 10 keV\@.  
In the Suzaku case, we found 5 sources whose fluxes were 
higher than $8\times10^{-14}$ erg cm$^{-2}$ s$^{-1}$ in the energy range 2 -- 10 keV\@.


\begin{thebibliography}{00}
\bibitem[Akahori \& Yoshikawa(2010)]{akahori10} Akahori, T., \& Yoshikawa, K.\ 2010, \pasj, 62, 335 

\bibitem[Anders \& Grevesse(1989)]{anders89} Anders, E., \&
  Grevesse, N.\ 1989, \gca, 53, 197

\bibitem[Bautz et al.(2009)]{bautz09} Bautz, M.~W., et al.\  2009, \pasj, 61, 1117

\bibitem[Borgani et al.(2004)]{borgani04} Borgani, S., et al.\ 2004, \mnras, 348, 1078

\bibitem[Burns et al.(2010)]{burns10} Burns, J.~O., Skillman, 
S.~W., \& O'Shea, B.~W.\ 2010, \apj, 721, 1105 

\bibitem[Cavaliere et al.(2011)]{cavaliere10} Cavaliere, A., Lapi, A., \& Fusco-Femiano, R.\ 2011, \aap, 525, A110 

\bibitem[Cen \& Ostriker(2006)]{cen06} Cen, R., \& Ostriker, J.~P.\ 2006, \apj, 650, 560 

\bibitem[Croston et al.(2008)]{croston08} Croston, J.~H., et al.\ 2008, \aap, 487, 431 

\bibitem[De Grandi \& Molendi(2002)]{degrandi02} De Grandi, S., \&  Molendi, S.\ 2002, \apj, 567, 163

\bibitem[Dickey \& Lockman(1990)]{dickey90} Dickey, J.~M., \& Lockman,
  F.~J.\ 1990, \araa, 28, 215

\bibitem[Eisenstein et al.(2005)]{eisenstein05} Eisenstein, D.~J.,  et al.\ 2005, \apj, 633, 560 

\bibitem[Eckert et 
al.(2011)]{eckert11} Eckert, D., Molendi, S., Gastaldello, F., \& Rossetti, M.\ 2011, \aap, 529, A133 



\bibitem[Ettori et al.(2002)]{ettori02} Ettori, S., De Grandi, S., \& Molendi, S.\ 2002, \aap, 391, 841 

\bibitem[Ettori et al.(2004)]{ettori04} Ettori, S., et al.\ 
2004, \mnras, 354, 111 

\bibitem[Evrard et al.(1996)]{evrard96} Evrard, A. E., Metzler, C. A.,
  \& Navarro, J. F., \ 1996, \apj, 469, 494

\bibitem[Fabricant et al.(1980)]{fabricant80} Fabricant, D., Lecar,
  M., \& Gorenstein, P.\ 1980, \apj, 241, 552

\bibitem[Forman \& Jones(1982)]{forman82} Forman, W., \& Jones, C.\ 1982, \araa, 20, 547 

\bibitem[Finoguenov et al.(2010)]{finoguenov10} Finoguenov, A., 
Sarazin, C.~L., Nakazawa, K., Wik, D.~R., 
\& Clarke, T.~E.\ 2010, apj, 715, 1143 


\bibitem[Fujita et al.(2008)]{fujita08} Fujita, Y., Tawa, N.,  Hayashida, K., Takizawa, M., Matsumoto, H., Okabe, N., 
\& Reiprich, T.~H.\ 2008, \pasj, 60, 343 



\bibitem[George et al.(2008)]{george08} George,M.R.,Fabian, A. C.,
  Sanders,J. S., Young., A. J., and Russell, H. R., \ 2008, MNRAS,
  395, 657

\bibitem[Hayashida et al.(1989)]{hayashida89} Hayashida, K.,
  Inoue, H., Koyama, K., \ 1989, \pasj, 41, 1373

\bibitem[Henry \& Briel(1996)]{henry96} Henry, J.~P., \& Briel, U.~G.\ 1996, \apj, 472, 137 

\bibitem[Henry(2000)]{henry00} Henry, J.~P.\ 2000, \apj, 534, 565
\bibitem[Henry et al.(2009)]{henry09} Henry, J.~P., Evrard, 
A.~E., Hoekstra, H., Babul, A., \& Mahdavi, A.\ 2009, \apj, 691, 1307 

\bibitem[Hoshino et al.(2010)]{hoshino10} Hoshino, A., et al.\ 2010, \pasj, 62, 371 

\bibitem[Ishisaki et al.(2007)]{ishisaki07} Ishisaki, Y., et al.\  2007, \pasj, 59, 113

\bibitem[Kawaharada et al.(2010)]{kawaharada10} Kawaharada, M., et  al.\ 2010, \apj, 714, 423 

\bibitem[Kalberla et al.(2005)]{kalberla05} Kalberla, P.~M.~W., Burton, W.~B., Hartmann, D., et al.\ 2005, \aap, 440, 775 

\bibitem[Komatsu et al.(2011)]{komatsu11} Komatsu, E., et al.\ 
2011, \apjs, 192, 18 

\bibitem[Koyama et al.(2007)]{koyama07} Koyama, K., et al.\ 2007,  \pasj, 59, 23

\bibitem[Kushino et al.(2002)]{kushino02} Kushino, A., Ishisaki, Y.,
  Morita, U., Yamasaki, N.~Y., Ishida, M., Ohashi, T., \& Ueda, Y.\
  2002, \pasj, 54, 327

\bibitem[Markevitch(1998)]{markevitch98v2} Markevitch, M.\ 1998, \apj,  504, 27 

\bibitem[Markevitch et al.(1998)]{markevitch98} Markevitch, M.,
  Forman, W.~R., Sarazin, C.~L., \& Vikhlinin, A.\ 1998, \apj, 503, 77

\bibitem[Markevitch et al.(2000)]{markevitch00} Markevitch, M., et al.\ 2000, \apj, 541, 542 

\bibitem[Markevitch  \& Vikhlinin(2007)]{markevitch07} Markevitch, M., \& Vikhlinin, A.\ 2007, \physrep, 443, 1 

\bibitem[Mitsuda et al.(2010)]{mituda10} Mitsuda, K., et al.\ 2010, \procspie, 7732,  

\bibitem[Nagai et al.(2007)]{nagai07} Nagai, D., Kravtsov, 
A.~V., \& Vikhlinin, A.\ 2007, \apj, 668, 1 


\bibitem[Nagai(2011)]{nagai11} Nagai, D.\ and Lau, E. T. 2011, ApJL in
  press, arXiv:1103.0280

\bibitem[Okabe \& Umetsu(2008)]{okabe08} Okabe, N., \& Umetsu, K.\ 2008, \pasj, 60, 345 

\bibitem[Pratt et al.(2007)]{pratt07} Pratt, G.~W., B{\"o}hringer, H.,
  Croston, J.~H., Arnaud, M., Borgani, S., Finoguenov, A., \& Temple,
  R.~F.\ 2007, \aap, 461, 71

\bibitem[Pratt et al.(2010)]{pratt10}Pratt, G.~W. and {Arnaud}, M. and {Piffaretti}, R. and {B{\"o}hringer}, H. and 
	{Ponman}, T.~J. and {Croston}, J.~H. and {Voit}, G.~M. and {Borgani}, S. and 
	{Bower}, R.~G. 2010, \aap, 511A,85P
	
\bibitem[Piffaretti 
\& Valdarnini(2008)]{piffaretti08} Piffaretti, R., \& Valdarnini, R.\ 2008, \aap, 491, 71 
	
	
\bibitem[Reiprich et al.(2009)]{reiprich09} Reiprich, T.~H., et al.\
  2009, \aap, 501, 899

\bibitem[{{Rudd \& Nagai}(2009)}]{rudd09} {Rudd},D.H., \& {Nagai},D.,
  2009, \apj, 701, L16

\bibitem[Serlemitsos et al.(2007)]{serlemitsos07} Serlemitsos, P.~J.,
  et al.\ 2007, \pasj, 59, 9

\bibitem[Simionescu et al.(2011)]{simionescu11} Simionescu, A., et al.\ 2011, Science, 331, 1576


\bibitem[Takei et al.(2008)]{takei08} Takei, Y., et al.\ 2008, 
\apj, 680, 1049

\bibitem[Tamura et al.(2008)]{tamura08} Tamura, T., et al.\  2008, \pasj, 60, 695 

\bibitem[Tawa et al.(2008)]{tawa08} Tawa, N., et al.\ 2008, 
\pasj, 60, 11 

\bibitem[Vikhlinin et al.(2005)]{vikhlinin05} Vikhlinin, A.,
  Markevitch, M., Murray, S.~S., Jones, C., Forman, W., \& Van
  Speybroeck, L.\ 2005, \apj, 628, 655

\bibitem[Vikhlinin et al.(2006)]{vikhlinin06} Vikhlinin, A., Kravtsov,
  A., Forman, W., Jones, C., Markevitch, M., Murray, S.~S., \& Van
  Speybroeck, L.\ 2006, \apj, 640, 691

\bibitem[Voit et al.(2003)]{voit03} Voit, G.~M., Balogh, 
M.~L., Bower, R.~G., Lacey, C.~G., \& Bryan, G.~L.\ 2003, \apj, 593, 272 

\bibitem[{{Voit}(2005)}]{voit05} {Voit},G,M.,2005,RvMP,77,207
  
 \bibitem[Zhang et 
al.(2006)]{zhang06} Zhang, Y.-Y., B{\"o}hringer, H., Finoguenov, A., Ikebe, Y., Matsushita, K., Schuecker, P., Guzzo, L., \& Collins, C.~A.\ 2006, \aap, 456, 55 

\bibitem[Umetsu et al.(2009)]{umetsu09} Umetsu, K., et al.\  2009, \apj, 694, 1643 

\bibitem[Urban et al.(2011)]{urban11} Urban, O., Werner, N., 
Simionescu, A., Allen, S.~W., {B{\"o}hringer}, H.\ 2011, \mnras, 546 

\end{thebibliography}

\end{document}